\documentclass[smallcondensed]{svjour3}
\usepackage[english]{babel}
\usepackage{amsmath}
\usepackage{graphicx}
\usepackage{amssymb}

\journalname{Celest Mech Dyn Astron}

\begin{document}

\title{Chirikov Diffusion in the Asteroidal Three-Body Resonance $\left(5,-2,-2\right)$}

\titlerunning{Chirikov-Arnold in the $\left(5,-2,-2\right)$ resonance}

\author{F. Cachucho  \and P. M. Cincotta \and S. Ferraz-Mello}

\authorrunning{F. Cachucho et al.}

\institute{F. Cachucho \at
           \email{fernando.cachucho@gmail.com}
      \and P.M. Cincotta \at
           Facultad de Ciencias Astron\'{o}micas y Geof\'{i}sicas \\
           Universitas National de La Plata, La Plata, Argentina \\
           \email{pmc@fcaglp.unlp.edu.ar}
      \and S. Ferraz-Mello \at
           Instituto de Astronomia, Geof\'{i}sica e Ci\^{e}ncias Atmosf\'{e}ricas \\
           Universidade de S\~{a}o Paulo, S\~{a}o Paulo, Brazil \\
           \email{sylvio@astro.iag.usp.br}}

\date{Received: date / Accepted: date}

\maketitle

\abstract{The theory of diffusion in many-dimensional Hamiltonian system is
applied to asteroidal dynamics. The general formulations
developed by Chirikov is applied to the Nesvorn\'{y}-Morbidelli analytic
model of three-body (three-orbit) mean-motion resonances (Jupiter-Saturn-asteroid system). In particular, we investigate the diffusion \emph{along} and \emph{across} the separatrices of the $\left(5,-2,-2\right)$ resonance of the (490) Veritas asteroidal family and their relationship to diffusion in semi-major axis and eccentricity. The estimations of  diffusion were obtained using the Melnikov
integral, a Hadjidemetriou-type sympletic map and numerical integrations
for times up to $10^{8}$ years.}


\section{Introduction}

The application of chaotic dynamics concepts to asteroidal dynamics
led to the understanding of the main structural characteristics of
asteroids distribution within the solar system. It was verified that
chaotic region are generally devoid of larger asteroids while, in
contrast, regular regions exhibit a great number of them (see
for instance, Berry 1978, Wisdom 1982, Dermott and Murray 1983,
Hadjidemetriou and Ichtiaroglou 1984, Ferraz-Mello et al. 1997,
Tsiganis et al. 2002b, Kne\v{z}evi\'{c} 2004, Varvoglis 2004)
It was soon
accepted that chaos was related inevitably to instability, which may
be local or global. Subsequent
investigations searched for initial conditions leading to instabilities
in relatively short time. In many applications, the determination
of Lyapunov exponent on a grid of initial conditions was used to get
quantitative informations on stability. The inverse of the largest
Lyapunov exponent, called Lyapunov time, should be in some way linked
to the characteristic time for the onset of chaos (Morbidelli and
Froeschl\'{e}, 1996).

However, some investigations have shown that many asteroids exhibit
intermediary behavior between chaos and regularity. The first registered
case was the asteroid (522) Helga (Milani and Nobili, 1992). This
asteroid was in chaotic orbit with a Lyapunov time inch shorter than the age of the solar system, but
it exhibited a long period stability. No significant evolution was
observed in the orbital elements of (522) Helga for times up to one
thousand times its Lyapunov time. Since then, other asteroids have
been shown to have Lyapunov times much shorter than the stability
times unraveled by simulations (e.g., Trojans, cf. Milani 1993). Currently,
this behavior is known in literature as \textit{stable chaos} (Milani
et al. 1997, Tsiganis et al. 2002a, Tsiganis et al. 2002b).  Indeed, there is
strong evidence that local instability does not mean chaotic diffusion, in the sense that nothing can be
said about how much global or local integrals (or orbital elements)
could change in a chaotic domain, even when a linear stability analysis
shows rather short Lyapunov times (see Giordano and Cincotta, 2004, Cincotta and
Giordano 2008).

Nesvorn\'{y} and Morbidelli (1998, 1999) demonstrated that one source
of stable chaos is related with three-body (three-orbit) mean-motion
resonances (Jupiter-Saturn-asteroid system). They observed that asteroids in these
resonances exhibit a slow diffusion in eccentricity and inclination, but no diffusion in the semi-major axis.
According to the estimates of Nesvorn\'{y} and Morbidelli (1998), about
1500 among the first numbered asteroids are affected by three-body
mean-motion resonances.

The three-body mean-motion resonances are very narrow since they appear
at second order in planetary masses, their typical width being $\sim10^{-3}$
AU, but they are much more dense (in phase space) than standard two-body
mean-motion resonances of similar size. Nesvorn\'{y} and Morbidelli (1999)
developed a detailed model for the three-body mean-motion resonance
and presented analytical and numerical evidence that most of them
exhibit a highly chaotic dynamics (at moderate-to-low-eccentricities)
which may be explained in terms of an overlap of their associated
multiplets. By multiplet, we refer to all resonances for which the
time-derivative of the resonant angle, $\sigma_{p,p_{J},p_{S}}$,
satisfies\begin{equation}
\dot{\sigma}_{p,p_{J},p_{S}}=m_{J}\dot{\lambda}_{J}+m_{S}\dot{\lambda}_{S}+m\dot{\lambda}+p\dot{\varpi}+p_{J}\dot{\varpi}_{J}+p_{S}\dot{\varpi}_{S}\simeq0,\label{resonant_angle}\end{equation}
for given $\left(m_{J},m_{S},m\right)\in\mathbb{Z}^{3}/\{0\}$. In
(\ref{resonant_angle}) the $\lambda$'s and $\varpi$'s denote, as
usual, the mean longitudes and perihelion longitudes, respectively;
$\left(p,p_{J},p_{S}\right)\in\mathbb{Z}^{3}$ are integers such that
$\sum_{i}\left(m_{i}+p_{i}\right)=0$ for \textit{i} ranging over
three bodies (Jupiter, Saturn and the asteroid).

We will be dealing in this paper with the case $\left(m_{J},m_{S},m\right)=\left(5,-2,-2\right)$.
This three-body resonance seems to dominate the dynamics of, for instance,
the asteroids (3460) Ashkova, (2039) Payne-Gaposchkin and (490) Veritas
(see Nesvorn\'{y} and Morbidelli 1999). In the case of the first two of
those asteroids (with relatively large eccentricity, $\sim0.15-0.20$),
their behavior looks regular over comparatively long time-scales (typically
$\sim1-10\times10^{3}$ years) while in case of (490) Veritas (with
eccentricity, $\sim0.06$) its dynamics looks rather chaotic over
similar time-scales. The determination of the age of (490) Veritas
family has been the concern of some authors who studied stable chaos
(Milani and Farinella 1994, Kne\v{z}evi\'{c} 1999, Kne\v{z}evi\'{c} et al. 2002, Kne\v{z}evi\'{c} 2003, Kne\v{z}evi\'{c} et al. 2004, Tsiganis et al. 2007, Kne\v{z}evi\'{c} 2007, Novakovi\'{c} et al. 2009).

Herein, we investigate chaotic diffusion \textit{along} (and also
\textit{across}) the above mentioned three-body mean-motion resonance
by means of a classical diffusion approach. We use (partially)
the formulations given by Chirikov (1979). That formulation are developed
to study specifically Arnold diffusion or some kind of diffusion that
geometrically resembles it initially called Fast-Arnold diffusion by
Chirikov and Vecheslavov (1989, 1993), as well as the so called
modulational diffusion (Chirikov et. al, 1985).
However, although from the purely mathematical point of view several
restrictions should be imposed, there are many unsolved aspects
regarding general phase space diffusion (see for instance Lochak 1999,
Cincotta 2002, Cincotta and Giordano 2008).

The structure of the Hamiltonian used by Chirikov in his first formulation
is similar to the
Hamiltonians obtained with the perturbations theories of Celestial
Mechanics. In particular, the Hamiltonians of analytic models of the
three-body mean-motion resonances are directly adaptable,
with some restrictions, to Chirikov's formulations.

Let us mention that some progress has been done
in the study of Arnold diffusion, particularly when applied to simple
dynamical systems, like maps, the latest ones are for instance, the
works of Guzzo et. al (2009a),(2009b), Lega (2009). However
the link between strictly Arnold diffusion and general diffusion
in phase space is still an open matter. Indeed, Arnold diffusion requires
a rather small perturbation, when the measure of the regular component
of phase space is close to one. Thus, as far as we know, almost
all investigations regarding Arnold diffusion involves relatively
simple dynamical systems like quasi--integrable maps.
In more real systems, like the one investigated in this paper,
the scenario is much more complex in the sense that the domain of
the three body resonance is almost completely chaotic.

Finally, this work is justified by the fact that an application
of all those theories to real astronomical models is still needed.

In Sect. 2, we summarize the general problem of computation of the
diffusion rate \textit{along} the resonance and we discuss the limitations
and difficulties to follow Chirikov approach in case of this particular
three-body mean-motion resonance. Section 3 is devoted to the resonant
Hamiltonian (given in Nesvorn\'{y} and Morbidelli (1999)) and its application
to the $\left(5,-2,-2\right)$ resonance. In Sect. 4, we construct
the simplified (or two-resonance) and complete (or three-resonance)
numerical models used in our investigations. Moreover, informations
about the algorithms and initial conditions used for numerical integrations
and the procedure for estimation of the diffusion are also considered
in this section. In Sect. 5, we discuss the numerical results on
diffusion in the $\left(5,-2,-2\right)$ resonance. In this application
of the Chirikov theory, we are concerned with the role of the perturbing
resonances in the diffusion \textit{across} and \textit{along} the
$\left(5,-2,-2\right)$ resonance and their relationship to diffusion
in semi-major axis and eccentricity. Finally, in Sect. 6, we investigate
the behavior of the asymptotic diffusion decreasing the intensity
of the perturbations in the $\left(5,-2,-2\right)$ resonance. In
this case, we are interested in the study of the diffusion under the
action of an arbitrarily weak perturbation considering scenarios close
to that of the Arnold diffusion.

\section{Chirikov's Diffusion Theory}

In this section we give Chirikov's (1979) as well as Cincotta's (2002)
description of diffusion theory in phase space in order to provide a
self-consistent presentation of the subject. Since most of the results and
discussions given here are included in at least these two reviews, we just
address the basic theoretical aspects.

Let us consider a Hamiltonian system having several periodic perturbations
that can create resonances. The initial conditions are chosen such that
the system is in the domain of a main resonance, called \textit{guiding
resonance}. The term of perturbation corresponding to the guiding
resonance is separated from the others, which will be called of \textit{perturbing
resonances}. The Hamiltonian has the following form
\begin{equation}
H=H_{0}\left(\mathbf{I}\right)+\epsilon V_{G}\left(\mathbf{I}\right)
\cos\left(\mathbf{m}_{G}\cdot\boldsymbol{\theta}\right)+
\epsilon V\left(\mathbf{I},\boldsymbol{\theta}\right),\label{Hamilton_original}
\end{equation}
with
\begin{equation}
\epsilon V=\epsilon\sum_{\mathbf{m}\neq\mathbf{m}_{G}}
V_{\mathbf{m}}\left(\mathbf{I}\right)\cos\left(\mathbf{m}\cdot\boldsymbol{\theta}\right),
\label{perturbation_original}
\end{equation}
where $V_{G}$ and $\mathbf{\mathbf{m}}_{G}$ are, respectively, the
amplitude and resonant vector of the \textit{guiding resonance}, $V_{\mathbf{m}}$
and $\mathbf{m}$ are, respectively, the amplitude and resonant vectors
of perturbing resonances. Here $\left(\mathbf{I},\boldsymbol{\theta}\right)$
are the usual \textit{N}-dimensional action-angle coordinates for
the unperturbed Hamiltonian $H_{0}\left(N\geq3\right)$, the vectors
$\mathbf{m}_{G}$, $\mathbf{m}\in\mathbb{Z}^{N}/\{0\}$ and $V_{G}$,
$V_{\mathbf{m}}$ are real functions. The small parameter perturbation,
$\epsilon$, is a real number such that $\epsilon\ll 1$. The resonance
condition is fixed by
\begin{equation}
S\left(\mathbf{I}^{r}\right)\mathbf{=}
\mathbf{m}_{G}\cdot\mathbf{\boldsymbol{\omega}}\left(\mathbf{I}^{r}\right)=0.
\label{eq:resonance_condition}
\end{equation}
The surface \textbf{$S\left(\mathbf{I}^{r}\right)=0$} in the action
space, is called \textit{resonant surface}.

\subsection{Dynamics of the guiding resonance in the actions space}

Let us first consider the simple case of one single resonance, that
is, let us assume that all $V_{\mathbf{m}}=0$ for $\mathbf{m}\neq\mathbf{m}_{G}$,
and we chose initial conditions close to the separatrix of the guiding
resonance. In the $\boldsymbol{\omega}$-space, the resonance condition
$\mathbf{m}_{G}\cdot\boldsymbol{\omega}^{r}=0$ has a very simple
structure, just a $\left(N-1\right)$-dimensional plane the
normal of which is the resonant vector $\mathbf{m}_{G}$. In the \textbf{$\mathbf{I}$}-space,
$\mathbf{m}_{G}\cdot\boldsymbol{\omega}^{r}=0$ leads to the $\left(N-1\right)$-dimensional
resonant surface $S\left(\mathbf{I}^{r}\right)=0$, whose local normal
at the point $\mathbf{I}=\mathbf{I}^{r}$ is

\begin{equation}
\mathbf{n}^{r}=
\left(\frac{\partial}{\partial\mathbf{I}}
\left[\mathbf{m}_{G}\cdot
\boldsymbol{\omega}\left(\mathbf{I}\right)\right]\right)_{\mathbf{I=}\mathbf{I}^{r}}
\label{eq:normalvector}
\end{equation}

In addition, we consider the $\left(N-1\right)$-dimensional surface
$H_{0}\left(\mathbf{I}\right)=E$ (in $\mathbf{I}$-space) and, if
we suppose that $\boldsymbol{\omega}\left(\mathbf{I}^{r}\right)$
is an one-to-one application, we can also write
$\widetilde{H}_{0}(\boldsymbol{\omega})=
H_{0}\left(\mathbf{I}\left(\boldsymbol{\omega}\right)\right)=E$
(in $\boldsymbol{\omega}$-space).

The manifolds
defined by the intersection of both resonant and energy surfaces has, in general,
dimension $N-2.$ By definition, the frequency vector $\boldsymbol{\omega}$
is normal to the energy surface in $\mathbf{I}$-space, since it is
the $\mathbf{I}$-gradient of $H_{0}.$ The latter condition, together
with the resonance condition Eqn. (\ref{eq:resonance_condition}),
shows that the resonant vector $\mathbf{m}_{G}$ lies on a plane tangent
to the energy surface at $\mathbf{I}=\mathbf{I}^{r}$. Furthermore,
the equations of motion (only with $H_{0}$
and the guiding resonant term) show that $\mathbf{\dot{\mathbf{I}}}$ is
parallel to the constant vector $\mathbf{m}_{G}$. Thus the motion under a single
resonant perturbation lies on the tangent plane to the energy surface
at the point $\mathbf{I}=\mathbf{I}^{r}$ in the direction of the
resonant vector.

\subsection{Local change of basis}

Now, let us introduce a canonical transformation
$\left(\mathbf{I},\boldsymbol{\theta}\right)\rightarrow\left(\mathbf{p},\boldsymbol{\psi}\right)$
by means of a generating function
\begin{equation}
F\left(\mathbf{p},\boldsymbol{\theta}\right)=\sum\limits _{i=1}^{N}\left(I_{i}^{r}+
\sum\limits _{k=1}^{N}p_{k}\mu_{ki}\right)\theta_{i},\label{geratrix_function}
\end{equation}
where $\mu_{ik}$ is a $N\times N$ matrix with $\mu_{1i}=\left(\mathbf{m}_{G}\right)_{i}$.
The transform action equations are
\begin{equation}
I_{i}=I_{i}^{r}+\sum\limits _{k=1}^{N}p_{k}\mu_{ki};\qquad\psi_{k}=
\sum\limits _{\ell=1}^{N}\mu_{k\ell}\theta_{\ell}.
\label{transform_equations}
\end{equation}
The phases $\psi_{k},k=1,\ldots,N$ are supposed to be non degenerate,
i.e., $\tfrac{\partial H_{0}}{\partial I_{k}}\neq0$. As Cincotta (2002) has shown,
this transformation
should better be thought as a local change of basis rather than as
a local change of coordinates. The action vector whose components
are $\left(I_{j}-I_{j}^{r}\right)$ in the original basis $\left\{ u_{j},j=1,\ldots,N\right\} $,
has components $p_{j}$ in the new basis $\left\{ \mu_{j},j=1,\ldots,N\right\} $
constructed taking advantage of the particular geometry of resonances
in action space.

We choose, $\mathbf{\boldsymbol{\mu}}_{1}=\mathbf{m}_{1}\equiv\mathbf{m}_{G}$
and since the vector $\mathbf{m}_{G}$ is orthogonal to the frequency
vector $\boldsymbol{\mathbf{\omega}}^{r}$ (due to the resonance condition),
it seems natural to take $\mathbf{\boldsymbol{\mu}}_{2}=
\boldsymbol{\mathbf{\omega}}^{r}/\left|\boldsymbol{\mathbf{\omega}}^{r}\right|.$
The remaining vectors of the basis are $\boldsymbol{\mathbf{\mu}}_{k}=
\mathbf{e}_{k},k=3,\ldots,N,$
the vectors $\mathbf{e}_{k}$ are orthonormal to each other and to
$\boldsymbol{\mathbf{\mu}}_{2}.$ Let us define one of the $\mathbf{e}_{k}$,
say $\mathbf{e}_{s}$, orthogonal to the normal $\mathbf{n}^{r}$
to the guiding resonance surface. In general, all the vectors $\mathbf{e}_{k}$
will be orthogonal also to $\boldsymbol{\mathbf{\mu}}_{1}$, except
$\mathbf{e}_{s}$. In general, $\mathbf{e}_{s}$ will not be orthogonal
to $\mathbf{m}_{G}$.
Then, considering $N=3$ and since $\mathbf{p}=p_{i}\boldsymbol{\mathbf{\mu}}_{i},
i=1,\ldots,3,$
we can say that $p_{1}$  measures the deviations of the
actual motion from the resonant point across the guiding resonance
surface, $p_{3}$ measures the deviation from the resonant value
along the guiding resonance, while $p_{2}$ measures the
variations in the unperturbed energy.

For $N\ge3$ degrees of freedom the subspace of intersection of the
two surfaces leads to a manifold of $N-2$ dimensions. Following Chirikov (1979),
this subspace is called \textit{diffusion manifold}. The $N-2$ vectors
$\mathbf{e}_{k}$ locally span (at the resonant value) a tangent plane
to the diffusion manifold called \textit{the diffusion plane.} Then,
in the new basis, the action vector may be written as:
$\mathbf{p}=p_{1}\mathbf{m}_{G}+p_{2}\boldsymbol{\mathbf{\omega}}^{r}/
\left|\boldsymbol{\mathbf{\omega}}^{r}\right|+\mathbf{q}$,
where $\mathbf{q}$ is confined to the diffusion plane
$\mathbf{q}=\sum_{k}q_{k}\mathbf{e}_{k}$
with $q_{k}=p_{k}$ for $k=3,\ldots,N$.

We write now the Hamiltonian (\ref{Hamilton_original}) in terms of
the new components of the action. Expanding up to second order in
$p_{k}$, using the orthogonal properties of the new basis, recalling
that $\psi_{1}$ is the resonant phase and neglecting the constant
terms, we obtain for $\left(k,\ell\right)\neq\left(1,1\right)$
\begin{equation}
H\left(\mathbf{p},\boldsymbol{\mathbf{\psi}}\right)\approx
\frac{p_{1}^{2}}{2M_{G}}+\epsilon V_{G}\cos\psi_{1}+\left|\omega^{r}\right|p_{2}+
\sum\limits _{k=1}^{N}\sum\limits _{\ell=1}^{N}\frac{p_{k}p_{\ell}}{2M_{k\ell}}+
\epsilon V\left(\boldsymbol{\psi}\right),
\label{Hamilton_transformed}
\end{equation}
with
\begin{equation}
\frac{1}{M_{k\ell}}=
\sum\limits _{i=1}^{N}\sum\limits _{j=1}^{N}\mu_{ki}\frac{\partial\omega_{i}^{r}}{\partial I_{j}}
\mu_{\ell j},
\label{tensor_massa}
\end{equation}

\begin{equation}
\frac{1}{M_{G}}=\frac{1}{M_{11}}=m_{Gi}\frac{\partial\omega_{i}^{r}}{\partial I_{j}}m_{Gj};
\label{mass_tensor_guiding}
\end{equation}
where we have written $V_{G},V\left(\boldsymbol{\psi}\right)$ instead
of $V_{G}\left(\mathbf{p}\right),V\left(\mathbf{p},\mathbf{\boldsymbol{\psi}}\right)$.
These functions are evaluated at the point $\mathbf{I}=
\mathbf{I}^{\mathit{r}}\textrm{ or }\mathbf{p=0}$.

In absence of perturbation $\left(V=0\right)$, the components $p_{k},k=2,\ldots,N$
are integrals of motion, which we set equal to zero so that $\mathbf{I}^{r}$
is a point of the orbit. Then the Hamiltonian (\ref{Hamilton_transformed})
reduces to
\begin{equation}
H\left(\mathbf{p,}\boldsymbol{\mathbf{\psi}}\right)\approx
H_{1}\left(p_{1},\psi_{1}\right)+
\epsilon V\left(\boldsymbol{\psi}\right),
\label{Hamilton_reduced}
\end{equation}
where
\begin{equation}
H_{1}=\frac{p_{1}^{2}}{2M_{G}}+\epsilon V_{G}\cos\psi_{1}
\label{Hamilton_pendulum}
\end{equation}
is the resonant Hamiltonian associated to the guiding resonance. It
is a simple pendulum. Note that the stable equilibrium point of the
pendulum is $\psi_{1}=\pi$ if $M_{G}V_{G}>0$, or $\psi_{1}=0$ if
$M_{G}V_{G}<0$.

To transform the phase variables, we take into account that the dot
product is invariant under a change of basis. Recalling
that $\psi_{k}=\sum_{\ell}\mu_{k\ell}\theta_{\ell}$,
then if $\mathbf{\boldsymbol{\nu}}$ denotes the vector $\mathbf{m}$
in the new basis, we have: $\varphi_{\mathbf{m}}\equiv
\mathbf{m}\boldsymbol{\cdot\mathbf{\theta}}=\boldsymbol{\mathbf{\nu}\cdot\mathbf{\psi}}$,
where $m_{k}=\sum_{\ell}\nu_{\ell}\mu_{\ell k}$. As we can see, while
the $m_{k}$ are integers, the quantities $\nu_{k}$ are, in general,
non-integer numbers, due to the scaling of the phase variables.

\subsection{Changes due to perturbation}

As mentioned above, for $V=0$ the $p_{k}$ are integrals of motion
and since $H_{1}$ is also an integral, we have the full set of $N$
unperturbed integrals: $H_{1},p_{2},q_{k},k=3,\ldots,N.$ But if we
switch on the perturbation, these quantities will change with time.
This can be seen using the equations of motion for the Hamiltonian
(\ref{Hamilton_transformed}), where $\dot{\psi}_{j}=\partial H/\partial p_{j},j=1,\ldots,N$.
Performing derivatives and integrating, considering
that for $V=0$, $p_{\ell}\left(\ell\neq2\right)$ are constants and
$p_{1}=M_{G}\dot{\psi}_{1}-\sum_{\ell=2}^{N}\frac{M_{G}}{M_{1\ell}}p_{\ell}$,
we obtain\begin{equation}
\begin{array}{ccc}
\psi_{k}\left(t\right)=\left|\boldsymbol{\omega}^{r}\right|t\delta_{2k}+\sum\limits _{\ell=2}^{N}\left(\frac{1}{M_{k\ell}}-\frac{M_{G}}{M_{k\ell}M_{1\ell}}\right)p_{\ell}t+\frac{M_{G}}{M_{k1}}\psi_{1}\left(t\right)+\psi_{k0}, &  & k>1\end{array}\label{new_phase}\end{equation}
 where $\delta_{ij}$ is the Kronecker's delta and $\psi_{j0}$ is
a constant. To get $\varphi_{\mathbf{m}}\left(t\right)$, we evaluate
the dot product $\sum_{i}\nu_{i}\psi_{i}$\begin{equation}
\varphi_{\mathbf{m}}=\mathbf{m}\cdot\boldsymbol{\mathbf{\theta}}=\boldsymbol{\mathbf{\nu}}\cdot\boldsymbol{\mathbf{\psi}}=\xi_{\mathbf{m}}\psi_{1}\left(t\right)+\omega_{\mathbf{m}}t+\beta_{\mathbf{m}}+K_{\mathbf{m}},\label{new_argument}\end{equation}
 where\begin{equation}
\xi_{\mathbf{m}}=\sum\limits _{k=1}^{N}\nu_{k}\left(\mathbf{m}\right)\frac{M_{G}}{M_{k1}},\qquad\omega_{\mathbf{m}}=\mathbf{\mathbf{m}}\cdot\mathbf{\boldsymbol{\omega}}^{r},\label{constant_argument}\end{equation}
 and $\beta_{\mathbf{m}}$ is a constant and\begin{equation}
K_{\mathbf{m}}=\sum_{\ell=2}^{N}\nu_{k}\left(\mathbf{m}\right)\left(\frac{1}{M_{k\ell}}-\frac{M_{G}}{M_{k\ell}M_{1\ell}}\right)p_{\ell}t.\label{eq:constphase}\end{equation}
The second relation of (\ref{Hamilton_pendulum}) is obtained taking
into account that \begin{equation}
\mathbf{m}\cdot\boldsymbol{\mathbf{\omega}}^{r}=\left(\sum_{i}\nu_{i}\left(\mathbf{m}\right)\boldsymbol{\mathbf{\mu}}_{i}\right)\cdot\left(\boldsymbol{\mathbf{\mu}}_{2}\left|\boldsymbol{\mathbf{\omega}}^{r}\right|\right)\label{eq:prodmomegar}\end{equation}
and the fact that, since $\mathbf{\boldsymbol{\mu}}_{2}$ is orthogonal
to all $\boldsymbol{\mathbf{\mu}}_{i},i\neq2$ and $\mu_{2}\cdot\mu_{2}=1$,
the dot product only contributes to $i=2$. Then,\begin{equation}
\mathbf{\omega_{\mathbf{m}}=m}\cdot\boldsymbol{\mathbf{\omega}}^{r}=\nu_{2}\left(\mathbf{m}\right)\left|\mathbf{\boldsymbol{\omega}}^{r}\right|.\label{eq:freq_m}\end{equation}

We are now ready to compute the time variation of the unperturbed
integrals. From (\ref{Hamilton_reduced}) and (\ref{perturbation_original}),
for $\dot{p}_{k}=-\partial H/\partial\psi_{k},k\neq1$, we easily
find\begin{equation}
\dot{p}_{k}\left(t\right)\approx\epsilon\sum\limits _{\mathbf{m\neq\mathbf{m}}_{G}}\nu_{k}\left(\mathbf{m}\right)V_{\mathbf{m}}^{r}\sin\varphi_{\mathbf{m}}\left(t\right).\label{time_variation_pk}\end{equation}
where $V_{\mathbf{m}}^{r}=V_{\mathbf{m}}\left(\mathbf{I}^{r}\right).$
This equation holds for every component of the momentum $\mathbf{p}$,
except for $p_{1}$. Since $p_{1}$ is not an integral, we use $H_{1}$,
instead of $p_{1}$.

Chirikov (1979) calculated the total variation of $H_{1}$ with
the aim of constructing a whisker map to describe the Arnold diffusion.
However, instead of it, we prefer, in the study of three-body resonances,
to compute the evolution of the components of the momentum
$\mathbf{p}$ by means of numerical integrations or, alternatively,
by mean of a Hadjidemetriou-type sympletic map (see Sect. 4). However,
we use a variation of Chirikov's construction to obtain a theoretical
estimate of the slow diffusion. We then proceed and compute the
total variation of $p_{k}$. For details about the construction of
the whisker map we refer to Chirikov (1979, Sect. 7.3) (see also
Cincotta 2002 and the Appendix B of Ferraz-Mello 2007).

If $\epsilon$ is small enough, the phase space domains associated
with all resonances present in (\ref{perturbation_original}) do not
overlap. Then a standard procedure is to replace $\psi_{1}\left(t\right)$
and $\dot{\psi}_{1}\left(t\right)$ by the values on the unperturbed
separatrix and to solve analytically (\ref{time_variation_pk}). We
make first the integration of (\ref{time_variation_pk}) over a complete
trajectory inside the stochastic layer assuming that $\psi_{1}=\psi_{1}^{sx}$
and $K_{m}=0$. Indeed, as mentioned previously for $V=0$ the $p_{\ell}\left(\ell\neq1\right)$
are integrals of motion and the phases $\varphi_{\mathbf{m}}$ can
be estimated considering $p_{\ell}\left(\ell\neq1\right)=0$, such
that $K_{\mathbf{m}}=0$. Then, the total variations of the $p_{k}$'s
are given by\begin{equation}
\Delta p_{k}\left(t\right)\approx\epsilon\sum\limits _{\mathbf{m\neq\mathbf{m}}_{G}}\nu_{k}\left(\mathbf{m}\right)V_{\mathbf{m}}^{r}\int_{_{-\infty}}^{^{+\infty}}\sin\varphi_{\mathbf{m}}^{sx}\left(t\right)dt,\label{eq:varpkint}\end{equation}
where $\varphi_{\mathbf{m}}^{sx}\left(t\right)=\xi_{\mathbf{m}}\psi_{1}^{sx}\left(t\right)+\omega_{\mathbf{m}}t+\beta_{\mathbf{m}}$. The estimate of integral into (\ref{eq:varpkint}) is done considering the known solutions for the phase $\psi_{1}^{sx}\left(t\right)$ obtained near both branches of unperturbed separatrix of the pendulum $H_{1}$. More details about these calculations are given in the appendix of this paper. Here we only described the main steps and the final result for $\Delta p_{k}\left(t\right)$.

Chirikov shown that the contributions of integral in (\ref{eq:varpkint}) in both branches of separatrix are described in terms of the Melnikov integral with arguments \begin{equation}
\pm\left|\lambda_{\mathbf{m}}\right|=\pm\left|\frac{\omega_{\mathbf{m}}}{\Omega_{G}}\right|,\label{eq:lambda}
\end{equation}
where the double sign indicates the both separatrix branches and $\Omega_{G}$ is the proper frequency of the pendulum Hamiltonian $H_{1}$. In order to simplify the calculations, Chirikov considered only even perturbing resonances and the contribution of Melnikov integral with negative argument was neglected under the condition $\left|\lambda_{\mathbf{m}}\right|\gg1$. In contrast, the perturbations in the three-body mean-motion resonance model are non even and the arguments are small. Moreover, the asymmetry in the Nesvorn\'{y}-Morbidelli model implies that the time of permanence of the motion near each separatrix is different. Thus, we introduce the factor $R_{T}$ which takes into account the difference in the time of permanence of the motion in each separatrix branch. Hence, after some algebraic manipulations the Eqn. (\ref{eq:varpkint}) is rewritten as
\begin{equation}
\Delta p_{k}\approx\frac{\epsilon}{\Omega_{G}}\sum\limits _{\mathbf{m}\neq\mathbf{m}_{G}}\nu_{k}\left(\mathbf{m}\right)Q_{\mathbf{m}}\sin\varphi_{\mathbf{m}}^{0},\label{eq:totalvarpk}\end{equation}
with\begin{equation}
Q_{\mathbf{m}}=V_{\mathbf{m}}^{r}\left[R_{T}A_{2\left|\xi_{\mathbf{m}}\right|}\left(\left|\lambda_{\mathbf{m}}\right|\right)+\left(1-R_{T}\right)A_{2\left|\xi_{\mathbf{m}}\right|}\left(-\left|\lambda_{\mathbf{m}}\right|\right)\right],\label{Melnikov_integral1}\end{equation}
where $\varphi_{\mathbf{m}}^{0}=\varphi_{\mathbf{m}}^{sx}\left(t=t^{0}\right)$
with $\psi_{1}^{sx}\left(t=t^{0}\right)=\pi$. Equation (\ref{eq:totalvarpk}) is a theoretical
estimate for the total variation of the momenta $p_{k}$'s inside the stochastic layer around
of separatrix of the pendulum Hamiltonian $H_{1}$, and it is valid for non-even perturbation
and for small $\lambda_{\mathbf{m}}$. Estimations of the Melnikov integral,
$A_{2\left|\xi_{\mathbf{m}}\right|}\left(\left|\lambda_{\mathbf{m}}\right|\right)$,
in terms of ordinary function can be obtained from the values of
$\left|\lambda_{\mathbf{m}}\right|$ and $\left|\xi_{\mathbf{m}}\right|$.
On the other hand, the factor $R_{T}$ can be estimated from numerical experiments.

\subsection{\label{sub:thediffusionrate}The diffusion rate}

In Chirikov's theory of slow diffusion, each resonance has a role
in the dynamics of system. The main resonance, that is the guiding resonance,
defines the domain where diffusion occurs. The stronger perturbing resonance
is called \textit{layer
resonance}. That resonance perturbs the guiding resonance separatrix
and it generates the stochastic layer and its
properties (width, KS-entropy, etc.). Thus, the layer resonance controls
the dynamics $\textrm{\underline{across}}$ the stochastic layer.
The weaker perturbing resonances are called \textit{driving resonances}.
They perturb the stochastic layer and control the dynamics $\textrm{\underline{along}}$
the stochastic layer. Then, the driving resonances are responsible
for the drift along the stochastic layer, i.e., the slow diffusion.
We are interested in obtaining an analytical estimate for the slow
diffusion. To fulfill this task, we will estimate the diffusion in
the actions whose direction is given along the stochastic layer.

We introduced the slow diffusion tensor\begin{equation}
D_{ij}=\frac{\overline{\Delta p_{i}\left(t\right)\Delta p_{j}\left(t\right)}}{T_{a}}\qquad i,j=3,\ldots N,\label{Diffusion_tensor}\end{equation}
where $T_{a}=\ln\left(32e/w_{s}\right)/\Omega_{G}$ is the characteristic
time of the motion within the stochastic layer of the guiding resonance
(equal to half the period of libration or to one period of circulation
of $\psi_{1}$ near the separatrix) and the average in the numerator
is done over successive values of $\varphi_{\mathbf{m}}^{0}$. Here $w_{s}$
is the width of the stochastic layer given by\begin{equation}
w_{s}=-\frac{\left|\boldsymbol{\omega}^{r}\right|}{\Omega_{G}^{2}}\omega_{\mathbf{\mathbf{m}}_{L}}\frac{\nu_{\mathnormal{1}}\left(\mathbf{\mathbf{m}}_{L}\right)\nu_{\boldsymbol{2}}\left(\mathbf{\mathbf{m}}_{L}\right)}{\xi_{\mathbf{\mathbf{m}}_{L}}}Q_{\mathbf{m}_{L}}>0.\label{width_layer}\end{equation}
(see Sects. 6.2 and 7.3 of Chirikov 1979). In the last equation,
the subscript $L$ indicate the layer resonance.
The components of the diffusion tensor (\ref{Diffusion_tensor}) are estimated using the Eqn. (\ref{eq:totalvarpk}). Hence, because of dependence with the phase $\varphi_{\mathbf{m}_{{\it D}}}^{0}$, the average in (\ref{Diffusion_tensor}) depends: (1) of the correlation between successive values $\varphi_{\mathbf{m}_{{\it D}}}^{0}$
when the system approaches the edges of the layer; (2) of the possible
interferences of several driving resonances. However, the analysis done by Chirikov shown that the terms that contribute to the diffusion must have the same phase $\varphi_{\mathbf{m}_{{\it D}}}^{0}$ (see Sect. 7.5 of Chirikov 1979 and Cincotta 2002 for more details). Hence, using (\ref{eq:totalvarpk}) the diffusion tensor components in (\ref{Diffusion_tensor}) are described as
\begin{equation}
D_{ij}=\frac{\epsilon^{2}}{T_{a}\Omega_{G}^{2}}\sum\limits _{\mathbf{m}_{D}}\nu_{i}\left(\mathbf{m}_{D}\right)\nu_{j}\left(\mathbf{m}_{D}\right)Q_{\mathbf{m}_{D}}^{2}\overline{\sin^{2}\varphi_{\mathbf{m}_{D}}^{0}}.\label{eq:diffusion_mean}\end{equation}
Terms with different $\mathbf{m}_{D}$ are averaged out.

Now, there still remains the problem of estimating $\overline{\sin^{2}\varphi_{\mathbf{m}_{D}}^{0}}$.
To solve this problem we need to consider that the structure of the stochastic layer
affects the motion of the system. In fact, studies of the slow diffusion theories have shown
that the stochastic layer is formed by two different regions.
The first, more central, near the unperturbed
separatrix, is totally chaotic. The second, more external, near the
edge of the stochastic layer, includes domains of regular motion forming
stability islands. When the solution approaches the edge of the stochastic
layer, it could remain rather close to the neighborhood of those stability islands
for long times. This phenomenon, called stickiness, leads to
a reduction in the diffusion rate (for more details about
the stickiness phenomenon see the recent work of Sun and Zhou 2009 and references
therein). Thus, near stability islands some
correlations in the phases arise, which dominate
the motion across and along the stochastic layer. In this case, the
evolution of phases $\varphi_{\mathbf{m}_{L}}^{0}$ and $\varphi_{\mathbf{m}_{D}}^{0}$
cannot be random simultaneously, and their correlation decreases the
diffusion rate (see Chirikov 1979, Cincotta 2002).

In order to estimate the correlation between $\overline{\sin^{2}\varphi_{\mathbf{m}_{D}}^{0}}$
and $\overline{\sin^{2}\varphi_{\mathbf{m}_{L}}^{0}}$, we use the so called \textit{reduced stochasticity approximation}, introduced by Chirikov (1979) like an additional hypothesis. Hence, the theoretical rate of diffusion given by (\ref{eq:diffusion_mean}) may be now evaluated and has the form
\begin{equation}
D_{ij}=\frac{\epsilon^{2}}{2\Omega_{G}^{2}T_{a}}\sum\limits _{\mathbf{m}_{D}}R_{\mathbf{m}_{D}}\nu_{i}\left(\mathbf{m}_{D}\right)\nu_{j}\left(\mathbf{m}_{D}\right)Q_{\mathbf{m}_{D}}^{2}\qquad i,j=3,\ldots,N.\label{Chirikov_diffusion_tensor}
\end{equation}
The Eqn. (\ref{Chirikov_diffusion_tensor}) is an estimate for the theoretical diffusion inside the stochastic layer. The diffusion coefficient includes two parameters that reduce the diffusion rate: $R_{T}$due to non-even perturbations and $R_{\mathbf{m}_{D}}$ due to the reduced stochasticity approximation.
The expression given here for the diffusion tensor is different of that given by Chirikov because of
the introduction of the parameter $R_{T}$ and by the possibility of having a small argument in the
Melnikov integral. Moreover, we have considered that the reduction factor due the reduced
stochasticity approximation is different for each driving resonance,
while Chirikov considers the same value for all of them.

\section{Application to 3-body mean-motion resonance}

The Hamiltonian, in the extended phase space,  associated to a given $\left(m_{J},m_{S},m\right)$
resonance, in Delaunay action-angles variables, is

\begin{equation}
H=-\frac{1}{2L^{2}}+n_{J}\Lambda_{J}+n_{S}\Lambda_{S}+v_{J}\Pi_{J}+v_{S}\Pi_{S}+\mathcal{P}_{\text{sec}}+\mathcal{P}_{\text{res}},\label{Hamilton_Nesvorny}\end{equation}
where

\begin{equation}
\lambda,\varpi,\lambda_{J},\varpi_{J},\lambda_{S},\varpi_{S}\label{Delaunay_angles}\end{equation}
are the mean longitudes and longitudes of the perihelions of the asteroid,
Jupiter and Saturn, respectively, and \begin{equation}
L=\sqrt{a};\Pi=\sqrt{a}\left(\sqrt{1-e^{2}}-1\right),\Lambda_{J},\Pi_{J},\Lambda_{S},\Pi_{S}\label{Delaunay_actions}\end{equation}
are the actions conjugated to them. The frequencies $n_{J},v_{J},n_{S},v_{S}$
are the mean-motion and perihelion motions of Jupiter and Saturn,
respectively.

The first term in (\ref{Hamilton_Nesvorny}) describes the Keplerian
motion of the asteroid and the terms proportional to the planetary
actions extend the phase space to incorporate the motion of the angles
$\lambda,\varpi,\lambda_{J},\varpi_{J},\lambda_{S},\varpi_{S}$ in
the unperturbed Hamiltonian. Details concerning the derivation of
this Hamiltonian are given by Nesvorn\'{y} and Morbidelli (1999), whose
main results and formula are used in this paper. Note that this Hamiltonian
does not satisfy the convexity condition, however, this fact should not be
a restriction for the application of Chirikov's diffusion theory.

The perturbing function, following Nesvorn\'{y} and Morbidelli (1999),
has been splitted into its secular and resonant parts
\begin{equation}
\mathcal{P}_{\text{sec}}=\frac{\mu_{J}}{a_{J}}\sum\limits _{k_{J},k_{S},k,i_{J},i_{S},i}P_{\text{sec}}\left(\alpha_{\text{res}}\right)e^{k}e_{J}^{k_{J}}e_{S}^{k_{S}}\cos\left(i_{J}\varpi_{J}+i_{S}\varpi_{S}+i\varpi\right)\label{Secular_perturbation}\end{equation}
\begin{equation}
\mathcal{P}_{\text{res}}=\frac{\mu_{J}}{a_{J}}\sum\limits _{k_{J},k_{S},k,p_{J},p_{S},p}P_{\text{res}}\left(\alpha_{\text{res}}\right)e^{k}e_{J}^{k_{J}}e_{S}^{k_{S}}\cos\left(\sigma_{p,p_{J},p_{S}}\right)\label{Resonant_perturbation}\end{equation}
where, $\alpha_{\text{res}}=a_{\text{res}}/a_{J}$ is the semi-major
axis corresponding to the exact resonance, $\sigma_{p,p_{J},p_{S}}=m_{J}\lambda_{J}+m_{S}\lambda_{S}+m\lambda+p\varpi+p_{J}\varpi_{J}+p_{S}\varpi_{S},$
$\mu_{J}$ is Jupiter's mass, $e$, $e_{J}$, $e_{S}$ are the asteroid,
Jupiter and Saturn's eccentricities, respectively, and $P_{\text{sec}}\left(\alpha_{\text{res}}\right)$,
$P_{\text{res}}\left(\alpha_{\text{res}}\right)$ are given functions that are linear in Saturn's mass
(see bellow). The harmonic coefficients satisfy d'Alembert rules,
$i_{J}+i_{S}+i=0$, $m_{J}+m_{S}+m+p+p_{J}+p_{S}=0$ and the series
are truncated at some order in $\left|k_{J}\right|+\left|k_{S}\right|+\left|k\right|$,
$\left|i_{J}\right|+\left|i_{S}\right|+\left|i\right|$ and $\left|m_{J}\right|+\left|m_{S}\right|+\left|m\right|+\left|p\right|+\left|p_{J}\right|+\left|p_{S}\right|$.
Next, we reduce the secular part (\ref{Secular_perturbation}) to
the quadratic term in asteroid's eccentricity in order to break the degeneracy
of the unperturbed Hamiltonian, and introduce in (\ref{Hamilton_Nesvorny})
the new action-angle variables, $\left(\mathbf{I'},\boldsymbol{\mathbf{\theta}'}\right)$:

\begin{equation}
\mathbf{I'}\boldsymbol{=}\left(N,N_{J},N_{S},\Pi,\Pi_{J},\Pi_{S}\right)\qquad\textrm{(actions)}\label{eq:new_action}\end{equation}

\begin{equation}
\boldsymbol{\mathbf{\theta}'=}\left(\nu,\tilde{\nu}_{J},\tilde{\nu}_{S},\varpi,\varpi_{J},\varpi_{S}\right)\qquad\textrm{(angles)}\label{eq:new_angle}\end{equation}
 defined by\begin{equation}
\nu=m_{J}\lambda_{J}+m_{S}\lambda_{S}+m\lambda,\qquad\tilde{\nu}_{J}=\lambda_{J},\qquad\tilde{\nu}_{S}=\lambda_{S},\label{new_old_angles}\end{equation}
 and\begin{equation}
L=mN,\qquad\Lambda_{J}=m_{J}N+N_{J},\qquad\Lambda_{S}=m_{S}N+N_{S}.\label{new_old_actions}\end{equation}
The variables ($\Pi_{J},\varpi_{J},\Pi_{S},\varpi_{S}$) remain unchanged.
We recall that the resonant perturbation (\ref{Resonant_perturbation})
does not depend on $\tilde{\nu}_{J}$ and $\tilde{\nu}_{S}$ (so that
$N_{J},N_{S}$ are constant that we can take as equal to zero). Let
us write\begin{equation}
\mathbf{I}\boldsymbol{\equiv}\left(N,\Pi,\Pi_{J},\Pi_{S}\right),\qquad\boldsymbol{\mathbf{\theta}\equiv}\left(\nu,\varpi,\varpi_{J},\varpi_{S}\right).\label{variables_transformed}\end{equation}
 Eliminating the constant terms, the Hamiltonian (\ref{Hamilton_Nesvorny})
may be written\begin{equation}
H\left(\mathbf{I,}\boldsymbol{\mathbf{\theta}}\right)=H_{0}\left(\mathbf{I}\right)+\tilde{V}\left(\mathbf{I,}\boldsymbol{\mathbf{\theta}}\right),\label{Hamilton_reduced_Nesvorny}\end{equation}
 where\begin{equation}
H_{0}\left(\mathbf{I}\right)=-\frac{1}{2m^{2}N^{2}}-\beta_{0}\left(1+\frac{\Pi}{mN}\right)^{2}+\left(m_{J}n_{J}+m_{S}n_{S}\right)N+\nu_{J}\Pi_{J}+\nu_{S}\Pi_{S},\label{Hamilton_Nesvorny_transformed}\end{equation}
 is the unperturbed Hamiltonian and the perturbation is described
by \begin{equation}
\tilde{V}\left(\mathbf{I,}\boldsymbol{\mathbf{\theta}}\right)=\sum\limits _{\mathbf{m}}\beta_{\mathbf{m}}\left(\mathbf{I}\right)\cos\left(\mathbf{m}\boldsymbol{\cdot\mathbf{\theta}}\right),\label{Perturbation_transformed}\end{equation}
 with\begin{equation}
\mathbf{m}=\left(1,p,p_{J},p_{S}\right),\label{eq:resonantvector}\end{equation}
 and\begin{equation}
\beta_{\mathbf{m}}\left(\mathbf{I}\right)=\frac{\mu_{J}}{a_{J}}\sum\limits _{k_{J},k_{S},k}P_{\text{res}}\left(\alpha_{\text{res}}\right)e^{k}e_{J}^{k_{J}}e_{S}^{k_{S}}.\label{Coefficient_perturbation}\end{equation}
 Nesvorn\'{y} developed a procedure allowing to obtain the coefficients
(\ref{Coefficient_perturbation}) in terms of power series of the
asteroid eccentricity only. In the last column of Table \ref{tabvectors}
are the coefficients calculated by Nesvorn\'{y} for the guiding (\textit{G}),
layer (\textit{L}) and driving (\textit{D}) resonances used in our
numerical experiments.

\begin{figure}
\centering
  \includegraphics[scale=0.4]{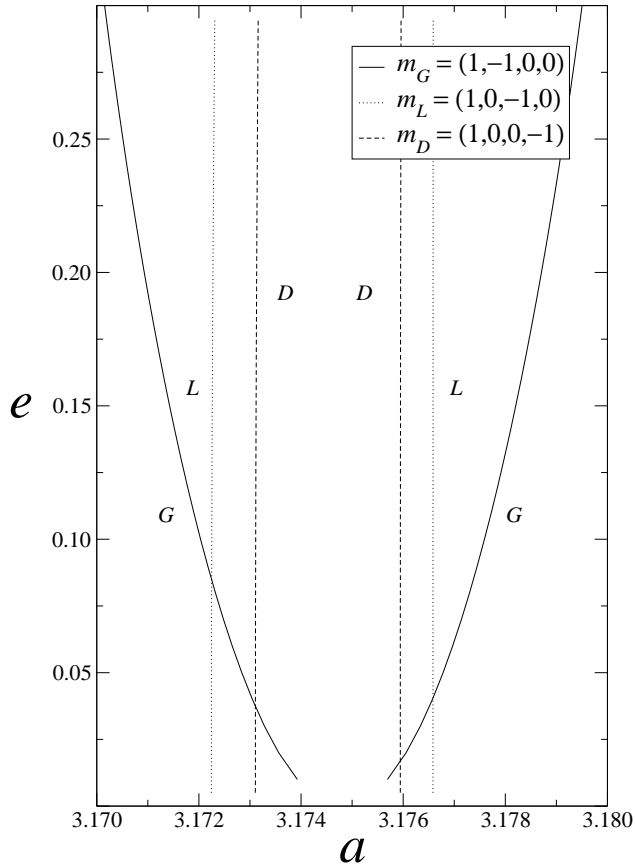}
  \caption{Unperturbed separatrices of guiding, layer and driving
resonances in the plane $(a,e)$. (Nesvorn\'{y} and Morbidelli, 1999)}
  \label{fig:figsep}
\end{figure}

We have considered the guiding resonance, defined by the vector $\mathbf{m}_{G}=\left(1,-1,0,0\right)$,
the layer resonance, defined by the vector $\mathbf{m}_{L}=\left(1,0,-1,0\right)$
and the driving resonance defined by vector $\mathbf{m}_{D}=\left(1,0,0,-1\right)$.
The unperturbed separatrices of those resonances in the plane $a-e$
are shown in Fig. (\ref{fig:figsep}).

The next step is to introduce the Chirikov variables $\left(\mathbf{p,}\boldsymbol{\psi}\right)$
allowing to have a separate representation of the actions \textit{across}
and \textit{along} the resonance within the stochastic domain of the
guiding resonance. The canonical transformation, is performed by the generating
function (\ref{geratrix_function}), with a transformation matrix,
$\mathbf{\boldsymbol{\mu}}$, given by

\begin{equation}
\boldsymbol{\mathbf{\mu}}  =  \left(\begin{array}{cccc}
1 & -1 & 0 & 0\\
\displaystyle{\frac{\omega_{2}^{r}}{\left|\boldsymbol{\omega}^{r}\right|}} & \displaystyle{\frac{\omega_{2}^{r}}{\left|\boldsymbol{\omega}^{r}\right|}} & \displaystyle{\frac{\nu_{J}}{\left|\boldsymbol{\omega}^{r}\right|}} & \displaystyle{\frac{\nu_{S}}{\left|\boldsymbol{\omega}^{r}\right|}}\\
\displaystyle{-\frac{2v_{S}n_{2}^{r}\omega_{2}^{r}}{\left|\mathbf{q}^{r}\right|}} & \displaystyle{\frac{2v_{S}n_{1}^{r}\omega_{2}^{r}}{\left|\mathbf{q}^{r}\right|}} & \displaystyle{-\frac{\nu_{J}v_{S}\left(n_{2}^{r}-n_{1}^{r}\right)}{\left|\mathbf{q}^{r}\right|}} & \displaystyle{\frac{\left|\mathbf{v}\right|^{2}\left(n_{2}^{r}-n_{1}^{r}\right)}{\left|\mathbf{q}^{r}\right|}}\\
\displaystyle{\frac{\sqrt{2}v_{J}}{2\left|\mathbf{v}\right|}} & \displaystyle{\frac{\sqrt{2}v_{J}}{2\left|\mathbf{v}\right|}} & \displaystyle{-\frac{\sqrt{2}\omega_{2}^{r}}{\left|\mathbf{v}\right|}} & 0\end{array}\right)\label{transformation_matrix}\end{equation}
where\[
\left|\mathbf{q}^{r}\right|=\sqrt{\left(v_{J}^{2}v_{S}^{2}+\left|\mathbf{v}\right|^{4}\right)\left(n_{2}^{r}-n_{1}^{r}\right)^{2}+4\nu_{S}^{2}\omega_{2}^{r2}\left|\mathbf{n}^{r}\right|^{2}},\]
 \[
\left|\mathbf{\mathbf{n}}^{r}\right|=\sqrt{\left(n_{1}^{r}\right)^{2}+\left(n_{2}^{r}\right)^{2}},\qquad\left|\mathbf{v}\right|=\sqrt{\nu_{J}^{2}+2\left(\omega_{2}^{r}\right)^{2}}.\]

\begin{table}
\caption{Old and new resonant vectors, and coefficients of the guiding (\textit{G}),
layer (\textit{L}) and driving (\textit{D}) resonances (The coefficients
were taken from Nesvorn\'{y} and Morbidelli, 1999).}              
\label{tabvectors}      
\centering                                      
\begin{tabular}{llll}          
\hline\noalign{\smallskip}
 &     vectors $\mathbf{m}$ &  vectors $\mathbf{\boldsymbol{\nu}}$ &  coefficients     $\widetilde{\beta}_{\mathbf{m}}\left(\times10^{-8}\right)$\tabularnewline
\noalign{\smallskip}\hline\noalign{\smallskip}
\textit{G } & $\left(1,-1,0,0\right)$ &  $(1,0,0,0)$ &  $45.59e-32.24e^{3}$\tabularnewline
\textit{L } &    $\left(1,0,-1,0\right)$ &  $(0.55,0.66,0.76,0.70)$ &  $-2.76+0.93e^{2}$\tabularnewline
\textit{D} & $\left(1,0,0,-1\right)$ &  $(0.68,0.47,0.92,-0.70)$ & $1.18-0.38e^{2}$\tabularnewline
\noalign{\smallskip}\hline                                            
\end{tabular}
\end{table}

Once the matrix of the transformation is defined, the new variables
$\left(\mathbf{p,}\boldsymbol{\mathbf{\psi}}\right)$ can be rapidly
obtained using the relations (\ref{transform_equations}). In the
new basis the arguments of the periodic terms change. The new vectors
$\mathbf{\boldsymbol{\nu}}$ defined by $\mathbf{m}\cdot\mathbf{\boldsymbol{\theta}}=\boldsymbol{\mathbf{\nu}}\cdot\boldsymbol{\mathbf{\psi}}$
are shown in Table \ref{tabvectors}, in addition to the resonant
vectors $\mathbf{m}$ and their respective coefficients.

The procedure of previous section was applied in the Nesvorn\'{y}-Morbidelli
model, and leads to the Hamiltonian (\ref{Hamilton_transformed})
with $N=4$. The three perturbation coefficients $\beta_{G}$, $\beta_{L}$
and $\beta_{D}$ of the guiding, layer and driving resonances, respectively,
are calculated at the resonant values $\mathbf{I}^{r}$, which satisfies
(\ref{eq:resonance_condition}). In the plane $N\Pi$ the resonant
condition (\ref{eq:resonance_condition}) leads to a curve satisfying
to\begin{equation}
\Pi^{r^{2}}+C_{1}\Pi^{r}+C_{2}=0,\label{eq:curvasresonantes}\end{equation}
 where $C_{1}\left(N^{r}\right)$ and $C_{2}\left(N^{r}\right)$ are
given in terms of $N^{r}$. Then, the solutions of (\ref{eq:curvasresonantes})
can be obtained analytically for a fixed value of $N^{r}$. However,
in Nesvorn\'{y}-Morbidelli model the coefficients $\tilde{\beta}_{\mathbf{m}}$'s
are given as functions of the asteroid eccentricity (see Table \ref{tabvectors}).
Therefore, we must use the definitions of Delaunay variables (\ref{Delaunay_actions})
to determinate $(a^{r},e^{r})$. The resonant eccentricity is determined
through $e^{r}=\sqrt{1-\left(1+\Pi^{r}/N^{r}\right)^{2}},$ where
$\left(N^{r},\Pi^{r}\right)$ satisfies (\ref{eq:resonance_condition}).
The resonant semi-major axis is determinate using $a^{r}=\left(N^{r}/2\right)^{2}.$

\section{Numerical Experiments}

In this section, we describe the numerical experiments done
to investigate the diffusion \textit{across} and \textit{along} the
stochastic layer of the three-body mean-motion resonance $\left(m_{J},m_{S},m\right)=(5,-2,-2)$ and its relations with the diffusion in semi-major axis and eccentricity.
In these investigations the diffusion \textit{across} will be described
by the actions $(p_{1},p_{2})$ and the diffusion \textit{along} by
the actions $(p_{3},p_{4})$. In order to determine the time evolution
of each action $\mathbf{p}$, we use the equations of motion obtained
from Hamiltonian (\ref{Hamilton_transformed}) with $N=4$. Then,
for each value $p_{k}(t)$, we use the equations of transformation
(\ref{transform_equations}) to obtain the respective values of $N(t)$
and $\Pi(t)$ and the definition of the Delaunay actions in (\ref{Delaunay_actions})
to obtain $a(t)$ and $e(t)$.

Two main models were considered in the numerical experiments: (i)
simplified (or two-resonance model) and (ii) complete (or three-resonance
model). In the first one, only one term of the perturbation - the
layer resonance - is considered. In the complete model, two terms
are considered: the layer and one driving resonance. In both cases
the guiding resonance is given by $\mathbf{m}_{G}=\left(1,-1,0,0\right)$.

Two different techniques were used to construct the solutions. In
a first set of experiments, the equations of motion of the Hamiltonian
(\ref{Hamilton_transformed}) were numerically integrated using the
Burlish-St\"{o}er method, for times in the interval $10^{2}\leq t_{int}\leq10^{8}$
years. The results of these simulations were sampled with an output
time step of 10 years. The simulations were done for the eccentricities
0.05 and 0.25 with the initial conditions given on the separatrix
of the guiding resonance $(p_{10}=2\sqrt{\left|M_{G}\beta_{G}^{r}\right|},p_{20}=0,p_{30}=0,p_{40}=0,\boldsymbol{\psi}=0)$.
The main goal in these experiments was the study of the variation
of the rate of diffusion \textit{across} and \textit{along} as a function
of the total time of the simulations. Moreover, we investigate the
correlations between the diffusion in the Chirikov actions $\mathbf{p}$
and the diffusion in semi-axis major and eccentricity.

In the other set of experiments, the simulations were done using an
Hadjidemetriou-type sympletic mapping (Hadjidemetriou 1986, 1988, 1991, 1993; Ferraz-Mello 1997; Roig and Ferraz-Mello 1999, Lhotka 2009) defined
by the canonical transformation $\left(\mathbf{p}^{n},\mathbf{\boldsymbol{\psi}}^{n}\right)\rightarrow\left(\mathbf{p}^{n+1},\mathbf{\boldsymbol{\psi}}^{n+1}\right),$
whose generating function is given by \begin{equation}
S\left(\mathbf{p}^{n+1},\mathbf{\boldsymbol{\psi}}^{n}\right)=\sum\limits _{i=1}^{3}\psi_{i}^{n}p_{i}^{n+1}+\eta H\left(\mathbf{p}^{n+1},\mathbf{\boldsymbol{\psi}}^{n}\right),\label{function_geratrix_map}\end{equation}
 where $\eta$ is the mapping step and the Hamiltonian is given by
(\ref{Hamilton_transformed}). The mapping equations are
\begin{equation}
p_{i}^{n+1}  =  p_{i}^{n}-\eta\frac{\partial H\left(\mathbf{p}^{n+1},\mathbf{\boldsymbol{\psi}}^{n}\right)}{\partial\psi_{i}^{n}}\label{equations_map}\end{equation}
\begin{equation}
\psi_{i}^{n+1}  =  \psi_{i}^{n}+\eta\frac{\partial H\left(\mathbf{p}^{n+1},\mathbf{\boldsymbol{\psi}}^{n}\right)}{\partial p_{i}^{n+1}}\qquad i=1,2,3.\label{eq:psi_map}
\end{equation}
 The procedure to determinate the semi-major axis and eccentricity
for each point $(p_{i}^{n+1},\psi_{i}^{n+1})$ of the trajectory is
analogous to that discussed above. The goal of these experiments is
to obtain the diffusion contour plots in the region of the 5,-2,-2
resonance in the plane $(a,e)$ (that plane is shown in Fig. \ref{fig:figsep})
for the two models considered. In this case, the total time of integration
used is the $10^{8}$ years with $\eta=10$ years. The initial conditions
are defined by the knots of a grid in the plane $(a,e)$, on the rectangle $\left(3.17\leq a_{0}\leq3.18\right)$U.A.,
$\left(0.01\leq e_{0}\leq0.30\right)$. The initial condition of the
state vector $\mathbf{p}_{0}$, for each point of the grid, was obtained
using the transformation equations (\ref{transform_equations}) and
the definitions of Delaunay variables. The initial condition for the
phases is $\psi_{k0}=0,k=1,\ldots4$. The use of the Hadjidemetriou
map was instrumental allowing the computation of the solutions starting on each point of the grid which, otherwise, would demand an excessively
large amount of CPU-time. The comparison of results provided by the map with those
obtained by integrating the Hamiltonian flow, do not show significant differences
in the numerical computation of the diffusion coefficient (see below), at least for the two values of the eccentricity used (0.05 and 0.25).

Finally, we need a numerical procedure to estimate the diffusion coefficient
of each element of the set $(p_{1},p_{2},p_{3},p_{4},a,e)$. In his
investigations, Chirikov (1979, et al. 1979, 1985) used a particular
method to determine the diffusion coefficient of the total energy
$H$ of the system. After Chirikov (1979), this procedure allows the
processes that are really stochastic to be separated from those associated
to bounded oscillations of periodic nature. Chirikov's procedure for
experimental determination of the diffusion coefficient consist in
dividing the total time of simulation $t_{int}$ in $N_{k}$ sub-intervals
of length $\left(\Delta t\right)_{k}$ and the calculation of the
mean value, $\bar{p}_{i}$, for every sub-interval. The contribution
to the diffusion rate for a given pair $\bar{p}_{i_{m}}$, separated
by interval of time $\left(m-\ell\right)\left(\Delta t\right)_{k}$,
is given by $\left(\bar{p}_{i_{m}}-\bar{p}_{i_{\ell}}\right)^{2}/\left|m-\ell\right|\left(\Delta t\right)_{k}$.
To obtain the rate of diffusion, the contributions of the considered
pairs are averaged over all the combinations $m\neq\ell$. That is,\begin{equation}
D_{i}^{k}=\frac{2}{N_{k}\left(N_{k}-1\right)}\sum_{m>\ell}\frac{\left(\bar{p}_{i_{m}}-\bar{p}_{i_{\ell}}\right)^{2}}{\left(\Delta t\right)_{k}\left(m-\ell\right)}.\label{diffusion_coef_exp}\end{equation}
The sub-intervals, used to estimate the mean values of quantities
$\bar{p}_{i}$, were obtained with $k=10$ and the length $(\Delta t)_{10}=t_{int}/10$.

The same procedure was used to determine the diffusion of the semi-major
axis, $D_{a}^{k}$, and of the eccentricity, $D_{e}^{k}$. We have
also estimated the eccentricity variation in these experiments using
a definition of diffusion rate of the random walking type (see for
example Eqn. (\ref{Diffusion_tensor})): \begin{equation}
\delta e\sim\sqrt{D_{e}^{k}t_{int}}.\label{variation_exp}\end{equation}

\section{Results and discussion}

In this sections, we discuss the results obtained in the numerical
experiments described above. In the discussion we will call action
\textit{across} to $(p_{1},p_{2})$, and \textit{across} diffusion
to $(D_{1},D_{2})$, where we suppressed the superscript $k$. In
the same way we call action \textit{along} to $(p_{3},p_{4})$ and
\textit{along} diffusion to $(D_{3},D_{4})$.

\subsection{\label{sub:role}The role of number of the perturbing resonances
in the diffusion}

In his theory, Chirikov showed that the number of perturbing resonances
is important for the dynamics of systems with many-dimensional Hamiltonians.
The results, in this case, repeat what is know from the general theory
of Hamiltonian systems. In a system with two degrees of freedom, the
resonances may be isolated by KAM tori, but for $(N>3)$ the dimensionality
may allows, in principle, a solution to visit the whole phase space when
$t\to\infty$.

\begin{figure*}
\centering
  \includegraphics[width=1.0\textwidth]{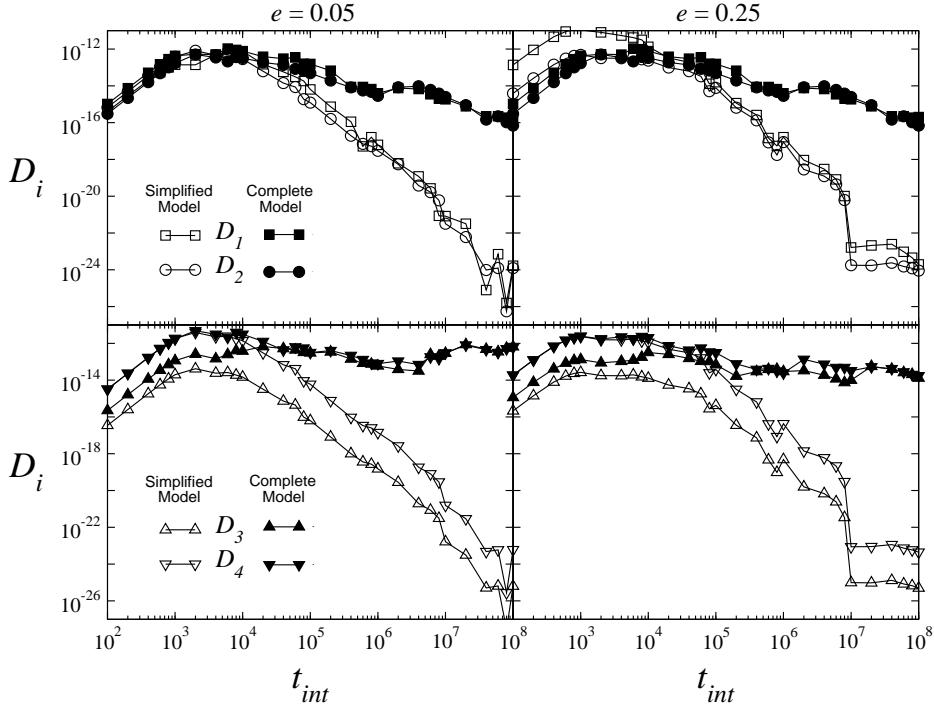}
  \caption{Diffusion coefficients of the actions associated
with motion \textit{across} and \textit{along} the guiding resonance,
in experiments over times from $10^{2}-10^{8}$\textit{ }years for
two different initial eccentricities.}
  \label{fig:resonumber}
\end{figure*}

Several experiments, using the Burlish-St\"{o}er integrator, were done
see the way in which the number of perturbing resonances in the diffusion
behavior. Figure \ref{fig:resonumber} shows the results for the diffusion
coefficients $D_{i},i=1,2,3,4$ in the simplified and complete models
as function of total integration time for eccentricities equal to
0.05 and 0.25, . In the plots of Fig. \ref{fig:resonumber}, we
see that the estimated diffusion increases in the low eccentricities
up to a maximum reached for $10^{3}-10^{4}$ years. This behavior
is explained by the fact that the solution needs to fill the stochastic
domain in the direction \textit{across} to it. After that maximum, in
the simplified model the diffusion coefficients for all actions decrease
continuously. This decrease indicates that the variation of the momenta
in both directions, \textit{across} and \textit{along} the stochastic
layer are bounded (as the total time increase, only the denominator
of (\ref{diffusion_coef_exp}) grows making the result to decrease).
As predicted by Chirikov's theory of slow diffusion, the actions $p_{3}$
and $p_{4}$ \textit{along} the resonance do not evolve, notwithstanding
the absence of topological barriers for its evolution. Without a driving
resonance, there is no long-period evolution of the solution \textit{along}
the stochastic domain. In our experiments, a very distinctive reduction
in the diffusion is observed in the case $e=0.25$ after $t_{int}\sim10^{7}$
years. This behavior is likely due to a sticking of the solution
to some regular domain.

The behavior of the diffusion in the complete model is more complicated.
In the experiment with $e=0.05$, the diffusion coefficients for the
actions \textit{across} the stochastic domain after $10^{8}$ years are
smaller than for the actions \textit{along} it. This difference reaches
approximately four orders of magnitude in this case and is due, probably,
to the limitation of the motion \textit{across} the stochastic layer
imposed by its width. For $e=0.25$ (right plot of Fig. \ref{fig:resonumber}),
the diffusion coefficients in the two models present almost the same
characteristics observed for $e=0.05$, except by the fact that, now,
the diffusion in the actions \textit{along} the stochastic layer,
present a slow reduction with the integration time after $10^{4}$
years. This behavior is likely due to the absence of overlapping of
resonance at high eccentricities, in contrast with the case of low
eccentricities, where the three resonances overlap (see Fig. \ref{fig:figsep}).

\subsection{Diffusion in semi-major axis and eccentricity in the complete model }

\begin{figure*}
\centering
  \includegraphics[width=1.0\textwidth]{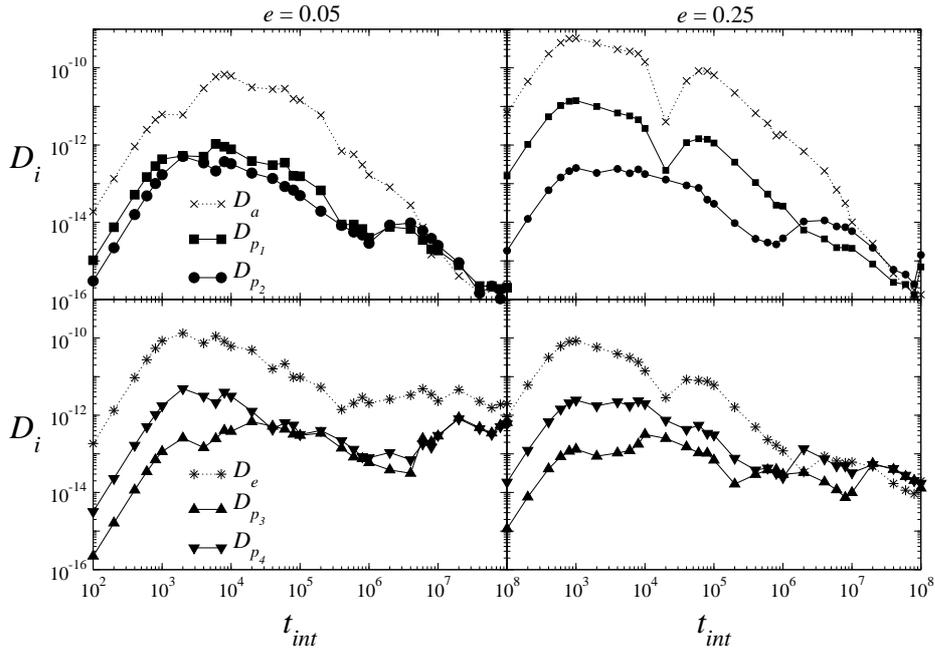}
  \caption{Diffusion Coefficient for actions across
and along, semi-major axis and eccentricity in experiments, obtained
for complete model, for times from $10^{2}$ up to $10^{8}$ years.
Each point is one experiment with initial conditions upon the unperturbed
separatrix of the guiding resonance.}
  \label{fig:orbitalelements}
\end{figure*}

The study of the previous section was completed
with the computation of the diffusion coefficients for the orbital
elements: semi-major axis and eccentricity in the complete model. Figure \ref{fig:orbitalelements} presents the results. The results for the actions shown in this figure are the same shown in Fig. \ref{fig:resonumber}, but with a magnified
scale. We see that, for large total times, there exist a correspondence
between the diffusion coefficients of the actions \textit{across}
$(p_{1},p_{2})$ and of the semi-major axis, and between the diffusion
coefficients of the actions \textit{along} the resonance $(p_{3},p_{4})$
and of the eccentricity. This behavior can be understood observing
the geometry of resonance $\left(5,-2,-2\right)$ shown in the Fig. \ref{fig:figsep}.
The separatrices of resonances are straight lines and the motion,
\textit{along} one of these separatrices, has constant semi-major
axis and variable eccentricity. Following the discussion presented in
Sect. \ref{sub:role}, and the comparison done in the previous section
for the simplified and completed models, we know that the drift \textit{along}
the separatrices only occurs if there is at least one driving resonance.
Hence, the eccentricity diffusion is due to the driving resonance.

As a complement to the previous discussion, we note that the variations
in semi-major axis occurs in the horizontal direction, the same direction
of the actions $(p_{1},p_{2})$. The behavior of the diffusion in
semi-major axis is similar to the diffusion of the actions \textit{across}
the resonance $(p_{1},p_{2})$ and is bounded by the width of the
stochastic domain. A consequence of this fact is that the diffusion
coefficient in the semi-major axis is smaller than that for the eccentricity
(in the complete model, the diffusion \textit{along} is not bounded).

\begin{figure*}
\centering
  \includegraphics[width=1.0\textwidth]{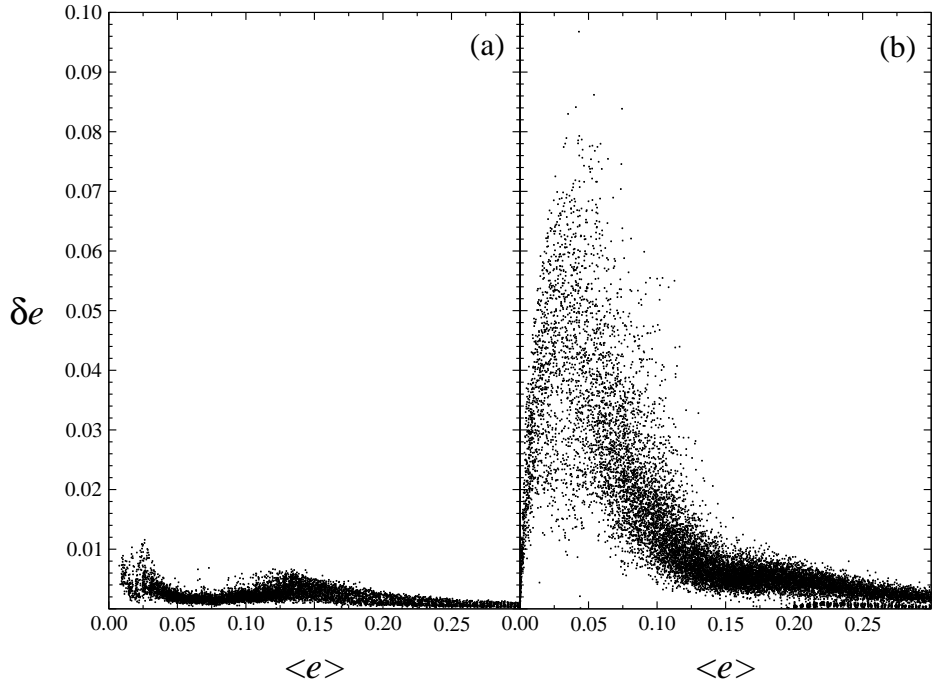}
  \vspace{2.5mm}
  \caption{Variation of the eccentricity versus
mean eccentricity for (a) simplified model and (b) complete model
on a net of points in the plane $(a,e)$.}
  \label{fig:eccentricityvariation}
\end{figure*}

The Fig. \ref{fig:eccentricityvariation} shows the variation of
the eccentricity calculated using initial conditions forming a grid
in the plane $(a,e)$ (the same grid of Fig. \ref{fig:contourn})
in $t_{int}=10^{8}$ years. The results for the simplified model are
shown in Fig. \ref{fig:eccentricityvariation}(a). In this case,
the larger variations in eccentricity occur for small values of the
eccentricity. Two shallow maximums are formed, which are likely related
with the eccentricity value at the intersection of the separatrices
of the guiding and layer resonances (the only secondary resonance
considered in the simplified model). The results for the complete
model are shown in Fig. \ref{fig:eccentricityvariation}(b). In
this case, the eccentricity variation reaches high values in the domain
of low eccentricities - between $0.01$and $0.125$ - with a maximum
for $<e>\sim0.05$. This maximum is certainly a result of the overlapping
of the resonances in low eccentricities, forcing the actions \textit{along}
the resonance.

In this model for mean eccentricities between 0.125 and 0.20 the variations
are of the same order. The distributions observed in the Fig. \ref{fig:eccentricityvariation}(b)
is in agreement with Nesvorn\'{y}'s unpublished
data for 45 numbered asteroids of the $(5,-2,-2)$ resonance (see
the Table 2 in Nesvorn\'{y} and Morbidelli 1998). The use of the models with only one perturbing resonance does not allow to get the distribution of the eccentricity variation observed in Fig. \ref{fig:eccentricityvariation}(b).

\subsection{The stochastic domain in the plane (\textit{a},\textit{e}). Dependence
on the initial conditions}

\begin{figure}
\centering
\includegraphics[width=1.0\textwidth]{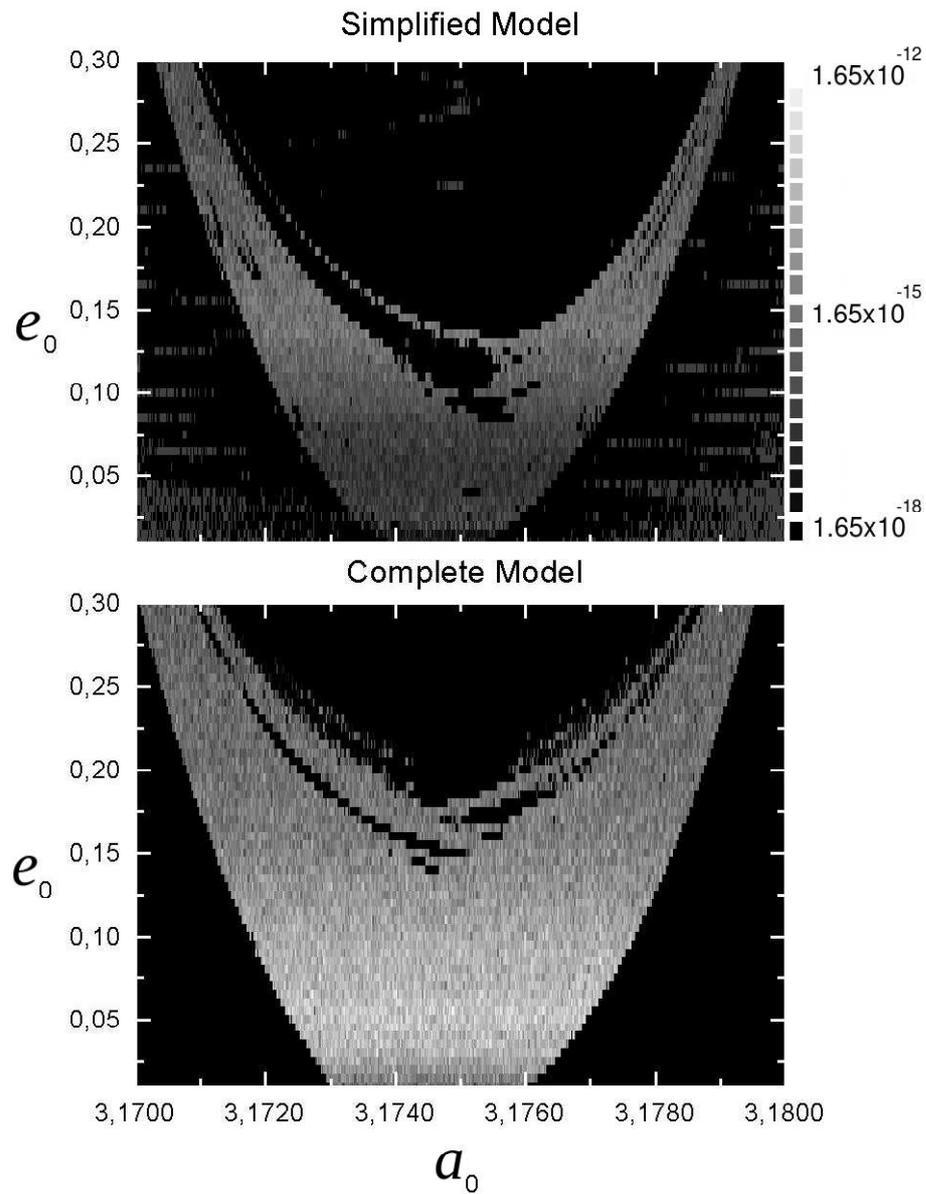}
  \caption{Diffusion coefficients for action \textit{along}
$p_{4}$ for initial conditions in the interval $3.17<a_{0}<3.18$
U.A. for semi-major axis and $0.01<e_{0}<0.30$ for eccentricity for
both simplified and complete models. The results were obtained for
total integration time equal to $10^{8}$ years.}
  \label{fig:contourn}
\end{figure}

The diffusion coefficients were calculated on a large set of initial
conditions to assess the domain where the solutions present stochastic
behavior. The analysis was done using simulations over $t_{int}\sim10^{8}$
years, on the points of a grid of initial conditions in plane $(a,e)$.
A Hadjidemetriou-type sympletic mapping was used instead of expensive
numerical integration to allow a large number of simulations. Figure
\ref{fig:contourn} shows the contour plots of the diffusion coefficients
of $p_{4}$. It shows the stochastic domain of the guiding resonance
(the light gray areas in Fig. \ref{fig:contourn}). Note that the
stochastic domain follows the geometry of the unperturbed separatrix
of Fig. \ref{fig:figsep}. Also note that the results for the complete
model show a stochastic domain $(a,e)$ larger than that observed for
the simplified model.

These differences are easily understood if we note the overlapping
of the three resonances in the considered range of eccentricities.
Figure \ref{fig:figsep} shows that the separatrices of the layer and
driving resonances are, for almost all eccentricities, interior to
the domain of the guiding resonance. At low eccentricities, however,
the separatrices cross one another. Thus, in low eccentricities, one
solution crossing the chaotic neighborhood of the separatrix of the
guiding resonance, also cross the separatrices of the layer and driving
resonances. The driving resonance acts pushing the actions \textit{along}
the guiding resonance. The magnitude of the push is determined by
the phase $\varphi_{\mathbf{m}_{D}}$ and amplitude $\beta_{\mathbf{m}_{D}}$.

At variance with the complete model, the simplified model presents
very low diffusion, in low eccentricities, as seen in Fig. \ref{fig:resonumber}.
In this case, the absence of the driving resonance (only the guiding
and layer are considered in the simplified model) implies in the
absence of evolution along the guiding resonance.

A remarkable feature in both results is the formation of a wide region,
in the central part of the domain of the guiding resonance, where
the diffusion is negligible. The motion appears regular for initial
conditions inside that region even when considering very long time
spans. This result confirms what Nesvorn\'{y} and Morbidelli (1999) observed
in surface of sections for eccentricity $0.20$ using this same analytic
model reduced to two degrees of freedom and two resonances. This is
different from the situation observed in low eccentricities, where
the separatrices of the resonances overlap.

\section{Asymptotic behavior}

\begin{figure*}
\centering
\includegraphics[width=1.0\textwidth]{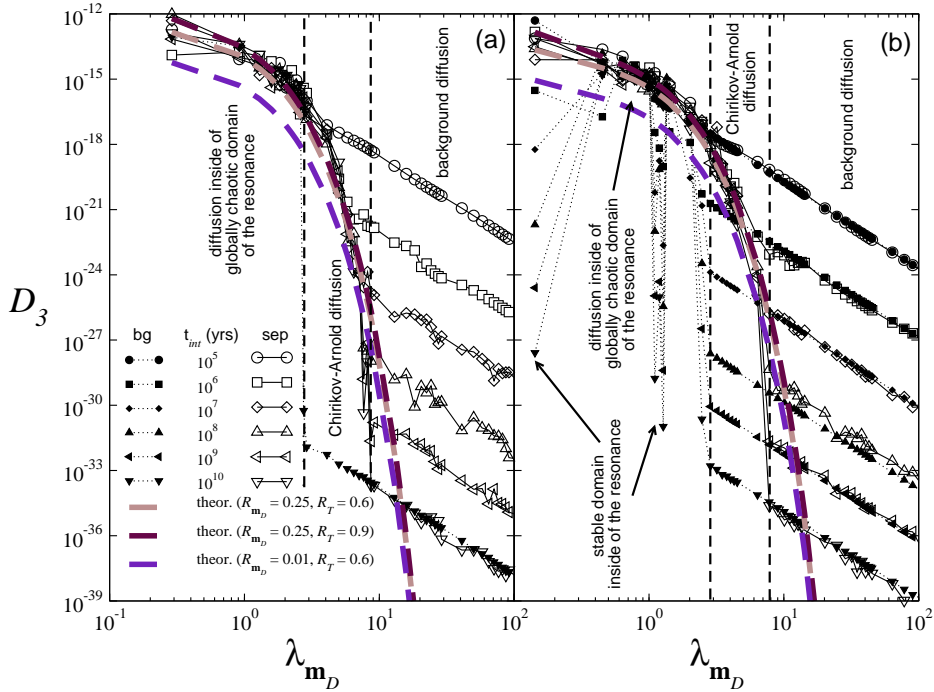}
  \caption{Asymptotic behavior of the diffusion coefficient $D_{3}$ for initial conditions over the separatrix and on the central part of the guiding resonance for (a) $e=0.05$ and (b) $e=0.20$. The dashed lines show the behavior predicted with Chirikov's theory.}
  \label{fig:asymptoticbehavior}
\end{figure*}

Chirikov theory of slow diffusion was constructed to study the diffusion
under the action of an arbitrarily weak perturbations, and the diffusion
coefficient was computed there using the asymptotic estimate of Melnikov's
Integral. The asymptotic behavior of the three-body resonance model
of Nesvorn\'{y} and Morbidelli was studied using the same technique devised
by Chirikov. Figure \ref{fig:asymptoticbehavior} shows the variation
of the diffusion coefficients \textit{along} the resonance, for two
different initial eccentricities (0.05 and 0.2), as functions of the
parameter $\lambda_{\mathbf{m}_{D}}=\omega_{\mathbf{m}_{D}}/\Omega_{G}$
appearing as argument of the Melnikov integrals in Sect. 2.3 in
the case the driving resonance $\mathbf{m}=\left(1,0,0,-1\right)$.
Small values of $\lambda_{\mathbf{m}_{D}}$ are obtained decreasing
the intensity of the guiding and perturbing resonances.

The Hadjidemetriou-like mapping was used to allow us to compute the solutions over $10^{10}$ years for a great deal of different conditions. The diffusion coefficient was calculated for initial conditions over
the separatrix of the guiding resonance. A background value $D_{b}$,
to be used as reference, was also obtained with initial conditions
in the central part of the guiding resonance, far from the separatrices.
For small values of the perturbation, the motion in the central part
of the resonance domain is regular and the background diffusion appear
as smaller than the diffusion shown by solution starting on the separatrices.
For high values of the perturbation intensity, the motion is chaotic
over the whole domain and the diffusion coefficients in the central
part are not different form those of solutions starting on the separatrices.

Figure \ref{fig:asymptoticbehavior} shows the diffusion coefficient
$D_{3}$ for $e=0.2$ and for the low eccentricity case $e=0.05$.
The figures for the coefficient $D_{4}$ are not shown since they
are almost identical to those shown for $D_{3}$. Figure \ref{fig:asymptoticbehavior}
shows the 3 different possibilities.

\begin{enumerate}
\item The first section of the figures, corresponding roughly to $\lambda_{\mathbf{m}_{D}}\lesssim2$,
is characterized by complete chaos. For the smallest $\lambda_{\mathbf{m}_{D}}$,
one sees the same phenomenon discussed in Sect. \ref{sub:role}:
the background diffusion appear small for some shorter runs because
they do not cover the time necessary to allow the solution to fill
the chaotic layer; but when time span grows, the diffusion values
increase as expected. In this section, in general the diffusion coefficients
for solutions starting in the central part or on the separatrices
are equal showing that the whole resonance domain is chaotic. A few
exceptions appear, as shown in Fig. \ref{fig:asymptoticbehavior}(b).
In addition we may see in this figure, for $\lambda_{\mathbf{m}_{D}}\sim1$,
a sudden decrease of the background diffusion indicating that the
corresponding solution stuck to some regularity island during its
evolution. However this sticking is not permanent and the background
diffusion grows when longer time spans are considered. The background
values are shown in Fig. \ref{fig:asymptoticbehavior}(a) only for
the time span $10^{10}$ years to allow a better comparison of the
numerical results with the dashed lines representing results from
Chirikov's model,
\item For $\lambda_{\mathbf{m}_{D}}\sim2$, the background diffusion shows
a discontinuity which, for the longest runs, reaches up to 14 orders
of magnitude. This means that the center of the resonance domain becomes
regular and the stochasticity remains confined to layers around the
separatrix. This is the domain where Chirikov's slow diffusion theories
are valid and where the results may be compared to the theoretical
results obtained in Sect. \ref{sub:thediffusionrate}. The
integration time is a crucial factor in the detection of the slow
diffusion. For instance, one may see that for simulations over only $10^{5}$ years, the diffusion near separatrix is equal to the background diffusion for values of $\lambda_{\mathbf{m}_{D}}$ close to 1, while for simulations over $10^{10}$ years, the equality is reached only for $\lambda_{\mathbf{m}_{D}}=9$.
\item In the last section of the Figs. \ref{fig:asymptoticbehavior} the
solutions starting close to the separatrices show a diffusion equal
to the background diffusion. The interpretations is that the stochastic
layer is this case is so thin that the used initial conditions are
no longer within them. (For that sake, the locus of the separatrices
should be computed with very large precision. See e.g. Froeschl\'{e} et
al. 2006). One striking feature in this section is that an increase
in the time span by a factor 10 means a decrease in the background
diffusion by a factor $10^{3}$. This is a clue for the fact that
the solutions are dominated by periodic terms. Indeed, if we consider
one periodic term with amplitude proportional to $\epsilon$ and frequency
$\omega$, its contribution to the average momentum in an interval
$\left[a,b\right]$ is proportional to \[
\frac{1}{\Delta t}\intop_{a}^{b}\epsilon\mbox{cos}\omega tdt\]
 where $\Delta t=b-a$. This integral is elementary and the integration
of the result over all frequencies below a upper limit $\omega_{\text{lim}}$,
gives \[
\frac{\epsilon}{\Delta t}\left[\mbox{si}\left(b\omega_{\text{lim}}\right)-\mbox{si}\left(a\omega_{\text{lim}}\right)\right]\]
where $\mbox{si}$ is the sine-integral function. The diffusion coefficients
are given by the square of the average variation of the momentum divided
by the total time (see Eqn. \ref{Diffusion_tensor}) and then $D\sim\Delta t^{-3}$.
We also have $D\varpropto\epsilon^{2}\sim\Omega_{G}^{-2}\sim\lambda_{D}^{-4}$.
The inclination -4 of the straight lines in the log-log plots can
be easily checked.
\end{enumerate}
The diffusion of the solutions in the neighborhood of the separatrix
may be determined from Eqn. \ref{Chirikov_diffusion_tensor}. This
equation involves the intensity of the perturbation (related to $\lambda_{\mathbf{m}_{D}}$)
and two unknown parameters: the factor of reduction $R_{\mathbf{m}_{D}}$
and the factor of odd perturbations $R_{T}$. The factor of reduction
corresponds to Chirikov's hypothesis of reduced stochasticity (due
to holes, the solution does not fill the strip around the separatrix);
the other factor comes from the fact that the perturbation is not
even and thus the values of the diffusion coefficient are not the
same for solutions in both separatrices (the solution may remain circulating
near one of the separatrices at time different of the time it remain
near the other).

The results obtained with Chirikov are shown in Figs. \ref{fig:asymptoticbehavior}
by dashed lines. In Figs. \ref{fig:asymptoticbehavior}(a) tree
different solutions are shown (calculated with the reduction factors
indicated in the figure). The better agreement is obtained with $R_{\mathbf{m}_{D}}=0.25$.
The two values used for $R_{T}$ (0.6 and 0.9) give almost the same
result, showing that the motion near deviation for the weakest perturbation
(larger $\lambda_{\mathbf{m}_{D}}$). In the other two figures, only
the two solutions with $R_{\mathbf{m}_{D}}=0.25$ are shown.

\section{Conclusion}

Chirikov's theories provide heuristic tools to understand the diffusion
observed in both eccentricity and semi-major axis of asteroids inside
the $\left(5,-2,-2\right)$ resonance. The multi-dimensional Hamiltonians of the
three-body (three orbit) mean-motion resonances may be studied with the
theories developed by Chirikov and collaborators,
mainly because of the particular geometry of those resonances in the
plane $\left(a,e\right)$. The results obtained
in this paper for the $\left(5,-2,-2\right)$ three-body mean-motion resonance
confirms the role of the resonances in the raising of diffusion \textit{across}
and \textit{along} the main resonance as foreseen in Chirikov's theories.

The diffusion calculations presented in this paper show that diffusion
in semi-major axis is related with the diffusion in the \textit{across}
actions $(p_{1},p_{2})$ while the diffusion in eccentricity is related
with the diffusion in the \textit{along} actions $(p_{3},p_{4})$.
The diffusion coefficient for the semi-major axis tends to small values
showing that the variation of the semi-major axis remains small. It
indicates the existence of barriers on both sides of the
stochastic layer limiting the motion \textit{across} the resonance.

The comparison between simplified and complete model results shown that the diffusion
in eccentricity is presumably due to the presence of at least one resonance driving
the motion along the guiding resonance. This behavior is similar to the expected behavior
of the Arnold diffusion, but, differently of it, the diffusion here is well apparent and
the diffusion coefficients remain high. For this reason it was sometimes
called Fast Arnold Diffusion (Chirikov and Vecheslavov 1989, 1993).

The structure of the $\left(5,-2,-2\right)$ resonance is formed by several overlapping
resonances, particularly at low eccentricities. Thus, diffusion across
$\left(5,-2,-2\right)$ resonance may be no longer limited to the thin chaotic
layers (stochastic layers), but it fills the whole resonance zone.
The diffusion \textit{along} the resonance could be due to a multiplet.
In this scenario, we have a random motion \textit{across} the resonance,
due to the overlap of several resonances belonging to a multiplet,
and another, likely due to weaker resonances, which drive the diffusion
\textit{along} the guiding resonance. Arnold diffusion
might occur inside the stochastic layer formed around the separatrix
of the guiding resonance under the action of sufficiently weak perturbations.
At variance, thick layer diffusion can appear for perturbation parameters
in a broad interval. Although this mechanism show a similar exponential
dependence of diffusion rate as a function of some system parameters,
the mean rate of thick layer diffusion is generally larger than any
theoretical estimation of Arnold diffusion. Therefore, it seems that for the
real problem would be more appropriate to call the diffusion with another name -
asymptotic diffusion of Chirikov-Arnold - due to the fact that this keeps some
features of the Arnold diffusion, but late very well characterized by
Chirikov like some distinct.

As we mentioned before, we believe that the connection between rigorous
investigations concerning strictly Arnold diffusion and that observed
in real physical systems like this, is still an open subject.
As Lochak (1999) pointed out, the global instability properties of near--integrable
Hamiltonian systems are far from well--understood.  It could almost be said
that little progress has been made after pioneering work Arnold,
and new ideas are definitely called for.

Finally, the good results showed that the Chirikov slow diffusion theory can be used
in broader investigations considering more resonances for $\left(5,-2,-2\right)$, as well applied for the
others three-body (three orbit) mean motion resonances and also can include the inclination
of the asteroid orbit.

\appendix
\section{Estimate of total variation of the momenta $p_{k}$'s}

To estimate the integral in (\ref{eq:varpkint} we use the approach done by Chirikov (1979). In fact, the unperturbed separatrix is defined by
\begin{equation}
\psi_{1}^{sx}\left(t\right)=4\textrm{arctan}\left(e^{\pm\Omega_{G}\left(t-t^{0}\right)}\right),\label{eq:psi1}\end{equation}
\begin{equation}
p_{1}^{sx}=\pm2\left|M_{G}\right|\Omega_{G}\textrm{sin}\frac{\psi_{1}^{sx}}{2},\label{eq:p1}\end{equation}
where \begin{equation}
\Omega_{G}=\sqrt{\epsilon\left|\frac{V_{G}}{M_{G}}\right|}\label{eq:freq_guiding}\end{equation}
is the proper frequency of the pendulum Hamiltonian $H_{1}$. The
double sign indicates the two separatrix branches: The positive sign
correspond to the upper separatrix $\left(0\leq\psi_{1}^{sx}<2\pi\right)$,
and the negative corresponds to the lower separatrix $\left(-2\pi\leq\psi_{1}^{sx}<0\right)$.
The stable equilibrium points lie at $\psi_{1}=\pm\pi$, respectively.
(This non usual separation of the intervals where the two branches
are considered allows $\psi_{1}$ and $p_{1}$ to have the same signal
in each separatrix and simplifies the next calculations. For the usual
presentation, the reader is referred to the study of the motions near
the separatrix of pendulum in the Appendix B of Ferraz-Mello, 2007.)

We introduce a time variable change $\tau=\Omega_{G}\left(t-t^{0}\right)$,
with $\psi_{1}^{sx}\left(t^{0}\right)=\psi_{1}^{0}=\pm\pi$. Then,
to $\psi_{1}^{s}>0$ (lower separatrix), we have\begin{equation}
\sin\varphi_{\mathbf{m}}^{sx}\left(t\right)=\sin\left(\xi_{\mathbf{m}}\psi_{1}^{sx}\left(\tau\right)+\lambda_{\mathbf{m}}\tau+\varphi_{\mathbf{m}}^{0}\right),\label{eq:phi_sx}\end{equation}
where $\varphi_{\mathbf{m}}^{0}=\xi_{\mathbf{m}}\psi_{1}^{0}+\omega_{\mathbf{m}}t^{0}+\beta_{\mathbf{m}}$,
with $\psi_{1}^{0}=\pi$, and \begin{equation}
\lambda_{\mathbf{m}}=\frac{\omega_{\mathbf{m}}}{\Omega_{G}}.\label{eq:lambda_m}\end{equation}
Or, after expansion of the right-hand side,\begin{equation}
\sin\varphi_{\mathbf{m}}^{sx}\left(t\right)  =  \sin\left(\xi_{\mathbf{m}}\psi_{1}^{sx}\left(\tau\right)+\lambda_{\mathbf{m}}\tau\right)\cos\varphi_{\mathbf{m}}^{0}+\cos\left(\xi_{\mathbf{m}}\psi_{1}^{sx}\left(\tau\right)+\lambda_{\mathbf{m}}\tau\right)\sin\varphi_{\mathbf{m}}^{0}.\label{eq:trig_rel}\end{equation}
When (\ref{eq:trig_rel}) is substituted into (\ref{eq:varpkint}),
the first term does not give contribution since, by symmetry, \begin{equation}
\int_{_{-\infty}}^{^{+\infty}}\sin\left[\xi_{\mathbf{m}}\psi_{1}^{sx}\left(\tau\right)+\lambda_{\mathbf{m}}\tau\right]d\tau=0.\label{eq:intnull}\end{equation}
The contribution of the second term of (\ref{eq:trig_rel}) is determined
by the relative signs of $\xi_{\mathbf{m}}$ and $\lambda_{\mathbf{m}}$.
Using the absolute values to $\xi_{\mathbf{m}}$ and $\lambda_{\mathbf{m}}$,
we can introduce the Melnikov integral in the form\begin{equation}
\int_{_{-\infty}}^{^{+\infty}}\cos\left(\left|\xi_{\mathbf{m}}\right|\psi_{1}^{sx}\left(\tau\right)\pm\left|\lambda_{\mathbf{m}}\right|\tau\right)d\tau=\frac{1}{\Omega_{G}}A_{2\left|\xi_{\mathbf{m}}\right|}\left(\mp\left|\lambda_{\mathbf{m}}\right|\right),\label{eq:cos_rel}\end{equation}
 where $A_{2\left|\xi_{\mathbf{m}}\right|}$ is the Melnikov integral
with argument $\pm\left|\lambda_{\mathbf{m}}\right|$. Then, the integral
in (\ref{eq:varpkint}) is\begin{equation}
\int_{_{-\infty}}^{^{+\infty}}\sin\varphi_{\mathbf{m}}^{sx}\left(t\right)dt=\frac{1}{\Omega_{G}}\textrm{sin$\varphi_{\mathbf{m}}^{0}$}A_{2\left|\xi_{\mathbf{m}}\right|}\left(\mp\left|\lambda_{\mathbf{m}}\right|\right).\label{eq:integralsine}\end{equation}

In the other branch of the separatrix, $\psi_{1}^{sx}$ has the signal
changed, but the particular symmetry of this equation makes it invariant
to the sign change of $\psi_{1}^{sx}$ and, thus, one obtains the same
result (\ref{eq:integralsine}). Indeed, if $\psi_{1}^{sx}<0$ the
parity of cosine makes the integral (\ref{eq:cos_rel}) to be \begin{equation}
\int_{_{-\infty}}^{^{+\infty}}\cos\left(\left|\xi_{\mathbf{m}}\right|\left|\psi_{1}^{sx}\left(\tau\right)\right|\mp\left|\lambda_{\mathbf{m}}\right|\tau\right)d\tau=\frac{1}{\Omega_{G}}A_{2\left|\xi_{\mathbf{m}}\right|}\left(\pm\left|\lambda_{\mathbf{m}}\right|\right),\label{eq:cos_rel_parity}\end{equation}
where we used $\psi_{1}^{sx}\left(\tau\right)=-\left|\psi_{1}^{sx}\left(\tau\right)\right|$.
Then, the variations in the actions $p_{k}$ can be obtained introducing
the result (\ref{eq:cos_rel_parity}) into (\ref{eq:varpkint}): \begin{equation}
\Delta p_{k}\left(t\right)\approx\frac{\epsilon}{\Omega_{G}}\sum\limits _{\mathbf{m\neq\mathbf{m}}_{G}}\nu_{k}\left(\mathbf{m}\right)V_{\mathbf{m}}^{r}\textrm{sin$\varphi_{\mathbf{m}}^{0}$}\left[A_{2\left|\xi_{\mathbf{m}}\right|}\left(\left|\lambda_{\mathbf{m}}\right|\right)+A_{2\left|\xi_{\mathbf{m}}\right|}\left(-\left|\lambda_{\mathbf{m}}\right|\right)\right].\label{eq:varpkchirikov}\end{equation}
Chirikov (1979) estimated the diffusion using the result of the last
equation. In order to simplify the theoretical estimate of the diffusion,
Chirikov considered only even perturbing resonances and neglected
the contribution of the perturbation with negative argument under
the condition $\left|\lambda_{\mathbf{m}}\right|\gg1$.

In the case of the three-body mean-motion resonance model, the perturbation
are non even and it is not possible to neglect the contribution of
perturbations for which $\left|\lambda_{\mathbf{m}}\right|$ is small.
Then, each perturbation contributes differently when the motion lies
close to a separatrix where $\lambda_{\mathbf{m}}>0$ or $\lambda_{\mathbf{m}}<0$.
Moreover, the odd perturbations in the Nesvorn\'{y}-Morbidelli model makes
necessary to take into account that the times of permanence of the
motion near each separatrix are not equal. This is done by considering
that the solution lies only a fraction of total time near the separatrix
with $\lambda_{\mathbf{m}}>0$. To take into account this asymmetry
we introduce the factor $R_{T}$\begin{equation}
R_{T}=\frac{T_{\lambda}}{T},\label{eq:factortime}\end{equation}
where $T_{\lambda}$ is the time that the solution stay in the neighborhood
of separatrix with $\lambda_{\mathbf{m}}>0$ and $T$ is the total
time. Then, the Eqn. (\ref{eq:varpkchirikov}) is rewritten as\begin{equation}
\Delta p_{k}\approx\frac{\epsilon}{\Omega_{G}}\sum\limits _{\mathbf{m}\neq\mathbf{m}_{G}}\nu_{k}\left(\mathbf{m}\right)Q_{\mathbf{m}}\sin\varphi_{\mathbf{m}}^{0},\label{eq:totalvarpkA}\end{equation}
with\begin{equation}
Q_{\mathbf{m}}=V_{\mathbf{m}}^{r}\left[R_{T}A_{2\left|\xi_{\mathbf{m}}\right|}\left(\left|\lambda_{\mathbf{m}}\right|\right)+\left(1-R_{T}\right)A_{2\left|\xi_{\mathbf{m}}\right|}\left(-\left|\lambda_{\mathbf{m}}\right|\right)\right].\label{Melnikov_integral}\end{equation}
Equation (\ref{eq:totalvarpkA}) is valid for non-even perturbation
and for small $\lambda_{\mathbf{m}}$. In order to obtain estimation
of (\ref{eq:totalvarpkA}) in terms of ordinary functions we must know
the values of $\left|\lambda_{\mathbf{m}}\right|$ and $\left|\xi_{\mathbf{m}}\right|$.
In general, the relations to $A_{2\left|\xi_{\mathbf{m}}\right|}$
depend on the exponential term with argument $\left|\lambda_{\mathbf{m}}\right|$
(see Appendix A in Chirikov 1979).

\section*{Acknowledgments}
The authors are grateful to an anonymous referee for a careful reading of the manuscript and helpful recommendations.
PMC is grateful to FAPESP (Brazil) for supporting his visit to the University of Sao Paulo.

\end{document}